\documentclass[sigconf]{acmart}
\AtBeginDocument{%
  }

\copyrightyear{2026}
\acmYear{2026}
\setcopyright{cc}
\setcctype{by}
\acmConference[SIGIR '26]{Proceedings of the 49th International ACM SIGIR Conference on Research and Development in Information Retrieval}{July 20--24,
  2026}{Melbourne, VIC, Australia.}
\acmBooktitle{Proceedings of the 49th International ACM SIGIR Conference on Research and Development in Information Retrieval (SIGIR '26), July 20--24, 2026, Melbourne, VIC, Australia}
\acmDOI{10.1145/3805712.3809683}
\acmISBN{979-8-4007-2599-9/2026/07}

\newtheorem{definition}{Definition}
\newtheorem{problem}{Problem}
\newtheorem{example}{Example}

\usepackage{multirow}
\usepackage{color}

\definecolor{lightgreen}{RGB}{50,150,50} 
\newcommand{\Qcolor}[1]{\textcolor{black}{#1}}
\newcommand{\QScolor}[1]{\textcolor{black}{#1}}

\begin{document}

\title{Multi-Faceted Continual Knowledge Graph Embedding for \\ \Qcolor{Semantic-Aware} Link Prediction}

\author{Jing Qi}
\orcid{0009-0008-7444-4057}
\affiliation{%
  \institution{Hangzhou Dianzi University}
  \city{Hangzhou}
  \country{China}
}
\email{jingqi@hdu.edu.cn}

\author{Yuxiang Wang}
\authornote{Corresponding authors.}
\affiliation{%
  \institution{Hangzhou Dianzi University}
  \city{Hangzhou}
  \country{China}}
\email{lsswyx@hdu.edu.cn}

\author{Zhiyuan Yu}
\affiliation{%
  \institution{Hangzhou Dianzi University}
  \city{Hangzhou}
  \country{China}
}
\email{zy\_yu@hdu.edu.cn}

\author{Xiaoliang Xu}
\affiliation{%
 \institution{Hangzhou Dianzi University}
 \city{Hangzhou}
 \country{China}}
\email{xxl@hdu.edu.cn}

\author{Yuanshi Zheng}
\affiliation{%
\institution{Xidian University}
  \city{Xi'an}
  \country{China}}
\email{zhengyuanshi2005@163.com}

\author{Tianxing Wu}
\authornotemark[1]
\affiliation{%
  \institution{Southeast University}
  \city{Nanjing}
  \country{China}}
\email{tianxingwu@seu.edu.cn}

\renewcommand{\shortauthors}{Jing Qi et al.}

\begin{abstract}
Continual Knowledge Graph Embedding (CKGE) aims to continually learn embeddings for new knowledge, i.e., entities and relations, while retaining previously acquired knowledge. Most existing CKGE methods mitigate catastrophic forgetting via regularization or replaying old knowledge. They conflate new and old knowledge of an entity within the same embedding space to seek a balance between them. However, entities inherently exhibit multi-faceted semantics that evolve dynamically as their relational contexts change over time. A shared embedding fails to capture and distinguish these temporal semantic variations, degrading lifelong link prediction accuracy across snapshots. To address this, we propose a \underline{M}ulti-\underline{F}aceted CKGE framework (MF-CKGE) for semantic-aware link prediction. 
During offline learning, MF-CKGE separates temporal old and new knowledge into distinct embedding spaces to prevent knowledge entanglement and employs semantic decoupling to reduce semantic redundancy, thereby improving space efficiency. 
During online inference, MF-CKGE adaptively identifies semantically query-relevant entity embeddings by quantifying their semantic importance, reducing interference from query-irrelevant noise. Experiments on eight datasets show that MF-CKGE achieves an average (maximum) improvement of 1.7\% (2.7\%) and 1.4\% (3.8\%) in MRR and Hits@10, respectively, over the best baseline. Our source code and datasets are available at: \url{https://anonymous.4open.science/r/MF-CKGE-04E5}.
\end{abstract}




\maketitle

\vspace{-0.3cm}
\section{Introduction}
Knowledge Graph Embedding (KGE) aims to preserve structural semantics within knowledge graphs (KGs) by projecting entities and relations into a low-dimensional vector space \cite{dong2014knowledge-KGs}.
The resulting embeddings are widely used in downstream tasks, including knowledge-enhanced RAG systems \cite{edge2024local-graphrag, li2024simple-subgraph, jimenez2024hipporag-hipporag}, recommendation systems \cite{zhang2016collaborative-recommender-systems}, 
\QScolor{question answering \cite{bordes2014open-questionAndanswering, WOS:001463224500001}, explainable argumentation \cite{Stathopoulos_Vassiliades_Diplaris_Vrochidis_Kompatsiaris_2024}}, and link prediction \cite{poursafaei2022towards,rossi2021knowledge-KGE3,zhang2018link}. Among these, link prediction, which predicts the tail (head) entity given a head (tail) entity and a relation, is often considered a fundamental task for evaluating the quality of KG embeddings. Only by obtaining high-quality KG embeddings can we further support advanced applications, \QScolor{such as recommendation systems \cite{tang2024editkg-link, xu2025rallrec-link, huang2023theory-link} and graph query \cite{9101747,9835709}}. Conventional KGE methods \cite{bordes2013translating-KGE1,sun2019rotate-RotatE,trouillon2016complex-Complex} operate under the assumption that the KG remains static during training. However, real-world KGs evolve dynamically, with new entities and relations continually emerging. Consequently, static embeddings demonstrate limited effectiveness in link prediction on evolving KGs. To cope with the growing nature, KGE methods typically resort to a retraining-from-scratch strategy. When updates are frequent and voluminous, this approach incurs substantial computational overhead. 
\QScolor{This motivates the research on continual learning methods over knowledge graphs \cite{chen2024continualmultimodalknowledgegraph},}  such as Continual Knowledge Graph Embedding (CKGE), which has attracted increasing attention in recent years.

\begin{figure}
    \centering
    \includegraphics[width=0.98\linewidth]{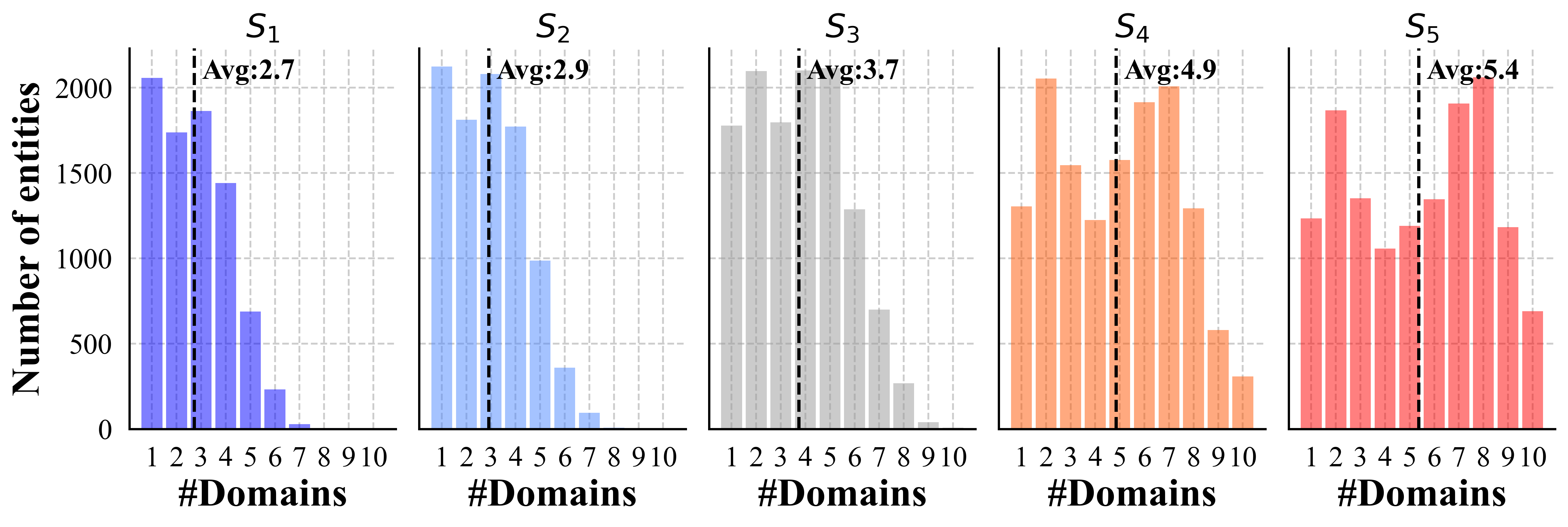}
    \vspace{-0.4cm}
    \caption{Distribution of knowledge domains for entities across evolving snapshots on the HYBRID dataset}
    \vspace{-0.8cm}
    \label{fig:trend}
\end{figure}

\vspace{0.1cm}
\noindent\textbf{Continual knowledge graph embedding.} Unlike KGE, CKGE continuously learns embeddings for new entities and relations while retaining the semantics of previously acquired knowledge, \Qcolor{thereby improving lifelong link prediction performance across evolving KG snapshots \cite{daruna2021continual-CKGE}.} Existing CKGE methods mainly focus on mitigating catastrophic forgetting, a phenomenon where models tend to overwrite or lose previously learned knowledge when adapting to new knowledge. Common strategies include replaying important triples~\cite{kou2020disentangle-DiCGRL, lopez2017gradient-GEM, wang2019sentence-EMR}, applying regularization mechanisms~\cite{cui2023lifelong-LKGE,liu2024towards-IncDE,zenke2017continual-SI}, and designing dynamic architecture \cite{liu2024fast-FastKGE,lomonaco2017core50-CWR,rusu2016progressive-PNN}. Despite differences in implementation, these methods typically unify the semantics of both new and old knowledge per entity into a shared embedding space for balanced representation. Although this strategy offers simplicity and practicality, it suffers from several limitations.

\vspace{0.1cm}
\noindent\textbf{Limitations.} In evolving KGs, entities often exhibit multi-faceted semantics depending on their contextual relations across different domains. For instance, a \texttt{Person}-type entity often embodies semantics from diverse domains such as family, career, and personal hobbies. As new knowledge (i.e., relations and entities) is incrementally added from heterogeneous sources, an entity’s context continually shifts, dynamically altering its semantics. We analyze the distribution of knowledge domains per entity across snapshots of evolving KGs. Figure \ref{fig:trend} illustrates the results on the HYBRID dataset \cite{cui2023lifelong-LKGE}. The $X$-axis represents the number of distinct knowledge domains per entity, and the $Y$-axis shows the number of such entities. Note that most entities are enriched with multi-faceted knowledge. \textbf{The average number of knowledge domains per entity rises steadily from 2.7 at snapshot $S_1$ to 5.4 at $S_5$}. Thus, maintaining a shared embedding for both new and old knowledge of an entity fails to capture these temporal semantic variations adequately, degrading link prediction accuracy across snapshots.

\begin{figure}
    \centering
    \includegraphics[width=0.94\linewidth]{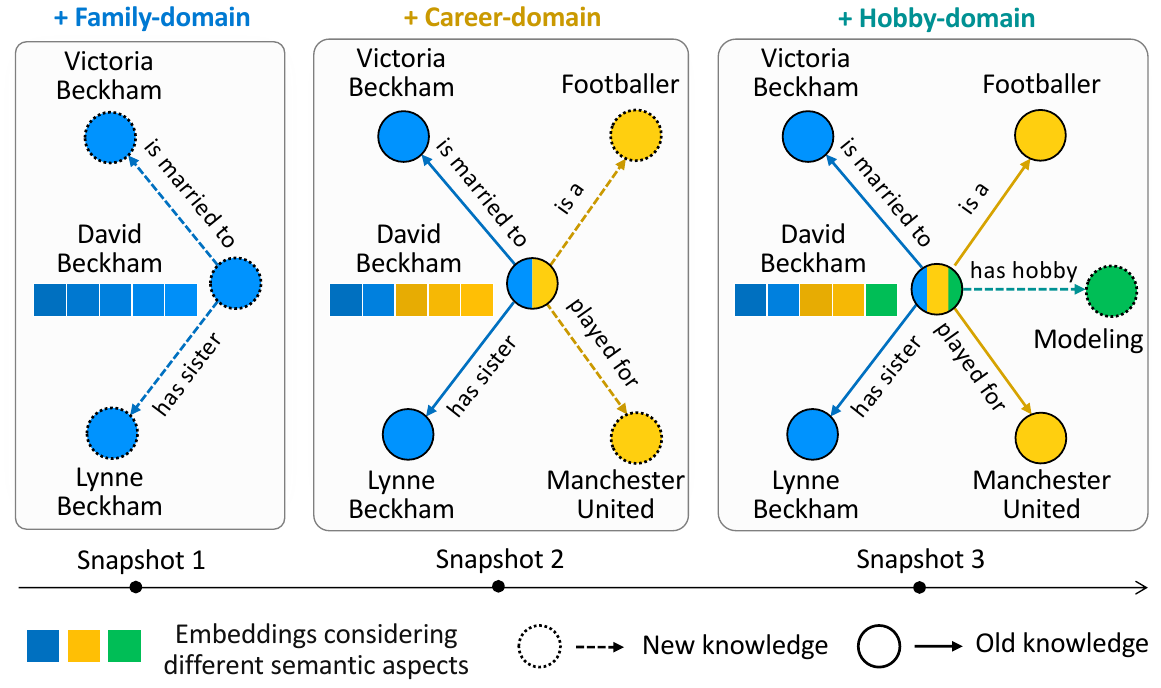}
    \vspace{-0.4cm}
    \caption{An evolving KG with three snapshots: each contains multi-faceted knowledge updates about \texttt{David Beckham}}
    \label{fig:example}
    \vspace{-0.7cm}
\end{figure}

Figure \ref{fig:example} illustrates three snapshots of an evolving KG. In snapshot 1, the context of \texttt{David Beckham} is composed of knowledge from his family-domain, reflecting the family-aspect semantics via a dedicated embedding vector (with each dimension in blue). As the KG evolves, new knowledge from his career-domain is added, making the single vector encode two distinct semantics (in blue and yellow). In snapshot 3, hobby-domain knowledge further expands this vector to represent three diverse semantics (in blue, yellow, and green). Clearly, using a fixed vector to represent \texttt{David Beckham}'s multi-faceted semantics across snapshots not only weakens the representational capacity as fewer dimensions are available per facet, but also causes semantic entanglement, which hinders link prediction. 
\QScolor{For example, when predicting \texttt{David Beckham}'s family-related tail entity at snapshot 3, 
only the family-related dimensions (blue region) are query-relevant. 
The remaining dimensions, which encode unrelated facets such as career or hobbies, introduce query-irrelevant noise and degrade prediction accuracy.}

\vspace{0.1cm}
\noindent\textbf{Our solution}. We propose a \underline{M}ulti-\underline{F}aceted CKGE (MF-CKGE) framework for semantic-aware link prediction, with two modules: (1) an offline \textit{entity embeddings decoupling} module (\S \ref{sec:decoupling}) that separates old and new knowledge in distinct embedding spaces to prevent knowledge entanglement, and (2) an online \textit{semantic-aware link prediction} module (\S \ref{sec:semantic_aware_link_prediction}) that adaptively adjusts the importance of each query-relevant embedding at inference time, suppressing irrelevant ones and improving prediction accuracy. For offline learning, we first distinguish between old and new knowledge via \textit{temporal decoupling}, that is, a dedicated embedding space is allocated for new knowledge, while retaining the existing space for old knowledge. Knowledge from different snapshots may originate from the same semantic domain, leading to semantic redundancy. We then employ a \textit{semantic decoupling} to identify and consolidate redundantly similar embeddings for the same entity across snapshots, while preserving only the most informative ones, thereby improving the space efficiency. During online inference, MF-CKGE quantifies the semantic relevance between the embeddings of each snapshot and the prediction query. These relevance scores then serve as runtime adaptive weights reflecting how important each snapshot embedding is for the query, to aggregate predictions derived from all snapshot embeddings, yielding the final result.

To sum up, the contributions of this paper are as follows:

\begin{itemize}
\item We propose the MF-CKGE framework, which captures multi-faceted semantics of entities and enables semantic-aware link prediction based on these diverse semantic aspects.
\item We propose an entity embeddings decoupling module in MF-CKGE that encodes new knowledge into distinct spaces to better differentiate it from old knowledge and reduce the semantic redundancy for space efficiency purposes.
\item We employ a semantic-aware link prediction module in MF-CKGE to automatically select semantically relevant entity embeddings to a specific query and adaptively adjust their importance in prediction, thereby enhancing the accuracy.
\item We conduct extensive experiments on eight datasets with diverse knowledge evolving patterns. The results show that MF-CKGE outperforms all baselines with an average 1.7\% and 1.4\% improvement in MRR and Hits@10, respectively.
\end{itemize}
\vspace{-0.3cm}
\section{Related Work}
Embedding learning for evolving KGs generally falls into two categories: Dynamic Knowledge Graph Embedding (DKGE) and Continual Knowledge Graph Embedding (CKGE).

\vspace{0.1cm}
\noindent\textbf{DKGE}. DKGE prioritizes efficiently synchronizing the embedding space with the current graph topology. It relies on partial retraining of global data to faithfully capture structural semantics in dynamic scenarios. puTransE \cite{tay2017non-puTransE} trains several models and performs non-parametric estimation of triplet scores to adapt to dynamic KGs; DKGE \cite{wu2022efficiently-DKGE} learns contextual embeddings for entities and relations to enhance the semantic accuracy of embeddings during incremental updates; 
iTransA \cite{jia2017knowledge-iTransA} supports the online optimization of entity-specific and relation-specific margins to incrementally update representations for graph growth.
However, they still require partial retraining on historical data, making it difficult to adapt to scenarios with significant knowledge growth. 
Therefore, to maintain the efficiency of incremental learning, it is necessary to eliminate the need for extensive retraining on historical data.

\begin{figure*}[t]
    \centering    \includegraphics[width=0.95\linewidth]{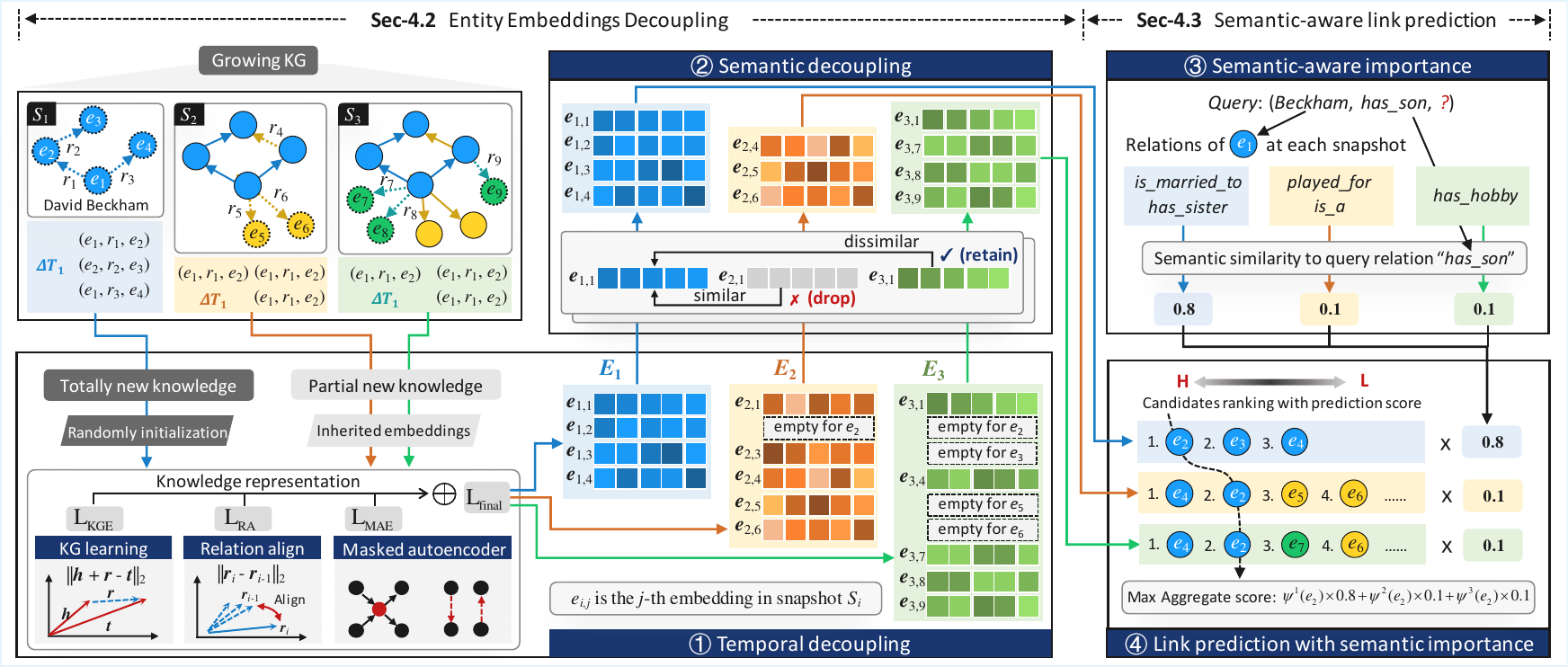}
    \vspace{-0.3cm}
    \caption{An overview of the proposed MF-CKGE framework}
    \label{fig:framework}
    \vspace{-0.5cm}
\end{figure*}

\vspace{0.1cm}
\noindent\textbf{CKGE}. CKGE aims to enable models to learn new knowledge from streaming updates while effectively preserving previously acquired information. Existing CKGE methods can be categorized into three main groups \cite{cui2023lifelong-LKGE}. (1) Dynamic-architecture-based methods \cite{lomonaco2017core50-CWR,rusu2016progressive-PNN} dynamically adapt their architecture to accommodate new knowledge without overwriting previously learned information.
For instance, FastKGE \cite{liu2024fast-FastKGE} handles updates through an incremental low-rank adapter (IncLoRA), which captures new knowledge while reducing update costs.
(2) Replay-based methods \cite{kou2020disentangle-DiCGRL,lopez2017gradient-GEM,wang2019sentence-EMR} replay a subset of old data when learning new knowledge, ensuring that the model retains previously learned knowledge. 
Notably, DiCGRL \cite{kou2020disentangle-DiCGRL} disentangles embeddings and handles updates by replaying relevant historical facts.
(3) Regularization-based methods \cite{cui2023lifelong-LKGE,kirkpatrick2017overcoming-EWC,zenke2017continual-SI} introduce regularization to constrain updates to critical parameters, ensuring stability. 
Specifically, IncDE \cite{liu2024towards-IncDE} employs incremental distillation as a regularization constraint to preserve structural knowledge.
All of these methods are designed to balance the acquisition of new knowledge with the retention of old knowledge within a single shared embedding space. However, this trade-off weakens the quality of representations for both new and previously learned knowledge, and these methods fail to suppress irrelevant information during link prediction. This motivates our study, which aims at eliminating knowledge entanglement and reducing the noise from query-irrelevant knowledge on link prediction.

\vspace{-0.1cm}
\section{Preliminaries}
\label{sec:preliminary}
Similar to prior studies \cite{cui2023lifelong-LKGE,liu2024towards-IncDE,liu2024fast-FastKGE}, we follow the standard CKGE settings, where the KG evolves through a sequence of incremental additions of entities and relations, defined as a growing KG.

\vspace{-0.05cm}
\begin{definition}[Growing Knowledge Graph]
We define the growing KG as a sequence of snapshots $G = \{S_1,S_2,\dots,S_n\}$. Each snapshot is a static KG denoted as a triplet $S_i =(E_i,R_i,T_i)$ at a specific time $t_i$, where $E_i$, $R_i$, and $T_i$ denote the entity, relation, and fact sets, respectively. Each fact is a triplet $(h,r,t) \in T_i$, where $h$, $t$ $\in$ $E_i$ represent the head and tail entities, and $r \in R_i$ is their relation. We use $\Delta E_i = E_i\setminus E_{i-1}$, $\Delta R_i = R_i \setminus R_{i-1}$, $\Delta T_i = T_i \setminus T_{i-1}$ to denote the newly added entities, relations, and facts during each update, respectively.
\end{definition}

\vspace{-0.05cm}
In this paper, we focus on designing a CKGE model for representation learning in growing KGs and validating its effectiveness through the \textit{lifelong link prediction task} \cite{cui2023lifelong-LKGE}. This task is a variant of classical link prediction, designed to measure the model's ability to learn new knowledge while retaining old knowledge.

\begin{problem}[Lifelong Link Prediction on Growing KGs]
Given a growing KG $G=\{S_1,S_2,\dots,S_n\}$, for each new fact set ${\Delta}T_{i}$ in the snapshot $S_i\in G$, it is divided into a training set $D_i$, a validation set $V_i$, and a test set $Q_i$. In the lifelong setting, it requires (1) using the $\langle D_1,V_1\rangle,\dots,\langle D_n,V_n\rangle$ in turn to train a CKGE model at each time $t_i$ ($i\in(1,n]$) to update the embeddings of existing entities $E_{i-1}$ and relations $R_{i-1}$, while obtaining the embeddings of new entities ${\Delta}E_i$ and relations ${\Delta}R_i$, and (2) evaluating the effectiveness of the CKGE model after finishing the learning on $\langle D_i,V_i\rangle$ via link prediction on the accumulated test data $\cup^i_{j=1}Q_j$, that is to predict missing head or tail entities given $(h,r,?)$ or $(?,r,t)$ from $\cup^i_{j=1}Q_j$ as queries.
\end{problem}
\vspace{-0.1cm}
Notably, once $\langle D_i,V_i\rangle$ is learned at time $t_i$, it is no longer available for subsequent learning. This design ensures that the CKGE model only learns new knowledge at time $t_i$. Besides, evaluating the model on the accumulated test data assesses both its degree of forgetting old knowledge and its capacity to learn new knowledge.

\vspace{-0.1cm}
\section{The Proposed MF-CKGE}
\subsection{Framework Overview}
The proposed MF-CKGE framework is illustrated in Figure \ref{fig:framework}. It consists of two main modules: \textit{entity embeddings decoupling} (\S \ref{sec:decoupling}) and \textit{semantic-aware link prediction} (\S \ref{sec:semantic_aware_link_prediction}). To alleviate the knowledge entanglement issue, the first module decouples entity embeddings from both the temporal and semantic perspectives: it separates old and new knowledge—arising as the KG evolves—through temporal decoupling, and mitigates semantic redundancy via semantic decoupling, thereby improving the model’s space efficiency. Subsequently, the second module adaptively adjusts the importance of query-relevant entity embeddings at runtime for on-the-fly link prediction queries, thereby improving link prediction accuracy by reducing the query-irrelevant noise with lower importance and enhancing query-relevant supportive facts with higher importance.

\subsection{Entity Embeddings Decoupling}
\label{sec:decoupling}

We decouple entity embeddings in two steps. The first step, called \textit{temporal decoupling}, operates from a temporal perspective: knowledge across different snapshots is categorized as either early-stage or recently arrived and is separated into distinct embedding spaces representing temporal old and new knowledge. The second step, termed \textit{semantic decoupling}, further decouples the entity embeddings from a semantic perspective. It identifies entities—spanning both temporal groups—whose embeddings exhibit highly similar semantics, merging these embeddings to eliminate semantic redundancy and retaining only the most informative representations.

\vspace{0.1cm}
\noindent\textbf{Temporal decoupling.}
Our primary objective is to capture the entity features specific to the current snapshot without being unduly constrained by historical data. To achieve this, a straightforward method is to freeze all embeddings learned in previous snapshots $\{S_1, \dots, S_{i-1}\}$, treating them as immutable temporal old knowledge. Subsequently, we focus on addressing the newly introduced knowledge within the current snapshot $S_i$ in two cases.

\vspace{0.1cm}
$\bullet$ \textbf{Case 1}: Totally new knowledge in $S_i$. We say a piece of knowledge is completely new if all three elements (head $h$, tail $t$, and relation $r$) of its factual triple $(h,r,t)$ are newly emerging in $S_i$.

\vspace{0.1cm}
$\bullet$ \textbf{Case 2}: Partially new knowledge in $S_i$. 
A piece of knowledge is partially new if its triple involves at least one new entity or relation, or is a novel triple formed among existing entities and relations.

\vspace{0.1cm}
In all cases, our goal is to obtain embeddings of new knowledge through representation learning of entities and relations (described in detail below). The key difference lies in how the initial embeddings of entities and relations are constructed under the two cases. For \textbf{totally new knowledge}, we apply random initialization for new entities and the relation in $(h,r,t)$, ensuring explicit isolation from historical representations. In contrast, for \textbf{partially new knowledge}, we initialize the existing entity or relation in $(h,r,t)$ by inheriting its embedding from the previous snapshot, while the newly introduced entity or relation remains randomly initialized. Notably, to minimize the training overhead, we restrict the input to the following representation learning to only those knowledge facts \QScolor{$\Delta D_i$} satisfying the above two cases, and exclude static knowledge, i.e., triples remaining unchanged in $S_i$, from being re-trained.
\QScolor{That is to say, if an existing entity does not appear in any of the new triples within $\Delta D_i$, it is completely excluded from the current snapshot's training process. Consequently, we neither perform inherited initialization nor allocate new embedding space for such entities.}

\vspace{0.1cm}
\noindent\underline{Entity and relation representation learning.} 
We design the loss function from three aspects to enable effective representation learning. \textbf{(1) Knowledge expressiveness}. This aspect focuses on encouraging the learned entity and relation embeddings to better capture the factual knowledge conveyed by a triple $(h,r,t)$. \textbf{(2) Cross-snapshot relation alignment}. Given that relation semantics in KGs are typically unique and stable \cite{wang2014knowledge-relation, ao2025lightprof-relation}, this aspect aims to maintain semantic consistency for embeddings of the same relation across different snapshots. \textbf{(3) Neighborhood semantic coherency}. This aspect requires semantic coherence between an entity's (or relation's) embedding and the one reconstructed from its neighborhood, ensuring that the semantics of the entity or relation remain logically aligned with its local context.

\vspace{0.1cm}
\noindent\textit{(1) Knowledge expressiveness}. We adopt the lightweight TransE \cite{bordes2013translating-KGE1}—a representative of the Trans-family KGE models—as the default implementation to strike a balance between efficiency and effectiveness. Alternatives such as widely used ComplEx \cite{trouillon2016complex-Complex} and RotatE \cite{sun2019rotate-RotatE} can also be deployed in our MF-CKGE. In our experiments, we evaluate their performance: MF-CKGE with TransE incurs the lowest time cost, while the others achieve higher effectiveness at the cost of longer training time. Regardless of the KGE model used, MF-CKGE consistently outperforms all baselines. Given the new knowledge $\Delta T_i$, the KGE loss is shown as Eq. \ref{eq:KGE}:
\begin{equation}
\label{eq:KGE}
\mathcal{L}_{\rm KGE} =\sum_{(h,r,t) \in \Delta T_i}max(0,f(\textbf{h}, \textbf{r}, \textbf{t}) - f(\textbf{h}', \textbf{r}, \textbf{t}') + \gamma)\ ,
\end{equation}

\vspace{-0.2cm}
\noindent where $(h', r, t')$ is a negative triple of $(h,r,t) \in \Delta T_i$, and $f(h,r,t) = \lVert \textbf{h} + \textbf{r} -\textbf{t}\rVert _p$ ($p=1$ or $2$, representing $L_1$ and $L_2$ norm) is the score function of TransE, with $\textbf{h}$, $\textbf{t}$, and $\textbf{r}$ denoting the embeddings of head entity, tail entity, and relation, from the learned entity and relation embedding spaces $\textbf{E}_i$ and $\textbf{R}_i$. Here, we use $L_1$ norm in $f(\cdot)$.

\vspace{0.1cm}
\noindent\textit{(2) Cross-snapshot relation alignment}. Unlike entities that exhibit multi-faceted semantics, a relation typically describes consistent semantics across snapshots. In order to reflect this, we employ a single embedding for each relation. 
Specifically, we apply a regularization term $\mathcal{L}_{\rm RA}$ during the training of snapshot $S_i$ to explicitly ensure the stability of existing relation embeddings as shown in Eq. \ref{eq:RA},
where $\textbf{r}_i$ and $\textbf{r}_{i-1}$ are embeddings of a relation $r\in R_i\cap R_{i-1}$ that appears simultaneously in both the current snapshot $S_i$ and the previous snapshot $S_{i-1}$. 
Consequently, only the most recent embedding $\textbf{r}_i$ is retained and used in training for the next snapshot.
\begin{equation}
\label{eq:RA}
\mathcal{L}_{\rm RA} =\sum_{r \in R_i\cap R_{i-1}}\lVert\textbf{r}_i - \textbf{r}_{i-1}\rVert_2^2
\end{equation}
\vspace{-0.2cm}

\vspace{0.1cm}
\noindent\textit{(3) Neighborhood semantic coherency}. In addition, to ensure semantic coherency between entities (or relations) and their local contexts, we adopt the \textit{masked autoencoder} (MAE) for self-supervised learning. Specifically, MAE utilizes a reconstruction-based strategy: for an entity $e$ (or relation $r$), it treats its first-order neighbors (or end-entities) as a masked local context; by aggregating the semantic features from local context following \cite{cui2023lifelong-LKGE}, it reconstructs the embeddings $\bar{\textbf{e}}$ (or $\bar{\textbf{r}}$). Then, the loss $\mathcal{L}_{\rm MAE}$ is defined as the distance between the original and reconstructed embeddings as follows.
\begin{equation}
\mathcal{L}_{\rm MAE} = \sum_{e \in E_i} \lVert\textbf{e} - \bar{\textbf{e}}\rVert_2^2  + \sum_{r \in R_i} \lVert\textbf{r}  - \bar{\textbf{r}}\rVert_2^2
\end{equation}

Finally, we combine all the aforementioned losses into the final loss $\mathcal{L}_{\rm final}$, where $\alpha$, $\eta$ are hyperparameters for balancing purposes.
\begin{equation}
\mathcal{L}_{\rm final} = \mathcal{L}_{\rm KGE} + \alpha\mathcal{L}_{\rm RA} + \eta\mathcal{L}_{\rm MAE}
\end{equation}

\noindent\textbf{Semantic decoupling.}
After temporal decoupling, we obtain a set of entity embeddings \(\mathbf{E}_i\) for each \(\Delta T_i\) (\(i\in[1,n]\)). To distinguish embeddings of the same entity across different snapshots, we denote the embedding of the \(j\)-th entity in the \(i\)-th snapshot as \(\mathbf{e}_{i,j}\). It is evident that the same entity may have similar semantics across various snapshots, and retaining all such embeddings would introduce semantic redundancy. For instance, a KG may incrementally add new facts about the same entity within the same domain at two snapshots, resulting in only minor changes to its embedding. To improve the spatial efficiency of embeddings, we introduce semantic decoupling. The key is to identify and consolidate the redundantly similar embeddings for the same entity across snapshots, while preserving only the most informative representations. Specifically, we examine all previous \QScolor{explicitly instantiated} embeddings $\textbf{e}_{x,j}$ at each snapshot $S_x$ ($x\in[1,i-1]$) and determine the one $\textbf{e}^*_{x,j}$ with the greatest semantic similarity to $\textbf{e}_{i,j}$ at the current snapshot $S_i$, where sim(·, ·) denotes the cosine similarity between two vectors.
\begin{equation}
\textbf{e}^*_{x,j} = \mathop{argmax}\limits_{\textbf{e}_{x,j}} \, sim(\textbf{e}_{x,j}, \textbf{e}_{i,j}), \forall x \in [1,i-1]
\end{equation}

Intuitively, if the semantic similarity between $\textbf{e}^*_{x,j}$ and $\textbf{e}_{i,j}$ exceeds a large enough threshold $\theta$, we consider them to be semantically redundant. To consolidate them, we adopt a lightweight yet effective strategy (Eq. \ref{eq:replace}): for $\textbf{e}_{i,j}$ with $sim(\textbf{e}^*_{x,j},\textbf{e}_{i,j})\geq \theta$, \textbf{we directly drop $\textbf{e}_{i,j}$ and retain only a pointer to $\textbf{e}^*_{x,j}$}. The primary motivation for discarding the new embedding $\textbf{e}_{i,j}$ rather than the historical $\textbf{e}^*_{x,j}$ is to maximize the retention of robust historical knowledge. 
\QScolor{During link prediction, this preserved pointer allows us to retrieve the most similar historical embedding $\textbf{e}^*_{x,j}$ and directly substitute it into the triple scoring function for inference.}
Conversely, $\textbf{e}_{i,j}$ with similarity \(<\theta\) is retained in \(\mathbf{E}_i\),
\QScolor{as such a divergence indicates that the entity has evolved a fundamentally new semantic facet; preserving it allows the model to expand its representational capacity to capture new knowledge domains.}
\begin{equation}
\label{eq:replace}
sim(\textbf{e}^*_{x,j},\textbf{e}_{i,j})\geq \theta\ ?\ \text{drop }\textbf{e}_{i,j}\ :\ \text{retain } \textbf{e}_{i,j}\ \text{in }\textbf{E}_i
\end{equation}

Here, $\theta$ serves as a semantic sensitivity parameter to regulate the degree of embedding consolidation. 
In our experiments, we evaluate its impact and show that it can effectively eliminate redundant embeddings while maintaining a good link prediction accuracy.
\QScolor{Furthermore, this pointer-based substitution naturally extends to the existing entities that are absent from the current incremental updates $\Delta D_i$. Since their semantics remain completely unchanged, we treat them as perfectly static. Given that no new embedding $\textbf{e}_{i,j}$ is generated for them during temporal decoupling, we directly assign a pointer referencing their most recent historical embedding. This strategy ensures evaluation consistency during inference without incurring any additional spatial overhead.}

\vspace{0.1cm}
\noindent\textbf{Remark.} While various consolidation strategies exist, such as mean, max, and sum pooling, they all replace snapshot-specific entity features with aggregated global information from multiple historical snapshots. Such aggregation operations again fuse embeddings from different snapshots, thereby reintroducing semantic entanglement to some extent, and resulting in performance degradation on downstream tasks. We have evaluated various embedding consolidation methods on link prediction tasks and demonstrated that our lightweight method effectively improves the space efficiency of embeddings while preserving model effectiveness. This suggests that retaining the integrity of robust historical embeddings is more beneficial than fusing them via aggregation.


\subsection{Semantic-Aware Link Prediction}
\label{sec:semantic_aware_link_prediction}
Given the decoupled entity embeddings $\{\textbf{E}_1,\dots,\textbf{E}_n\}$ for snapshots $\{S_1,\dots,S_n\}$, we now discuss how to leverage them for link prediction. For a specific prediction query $(h,r,?)$, a key insight is that not all historical embeddings of an entity are equally relevant to the query--each may capture a different semantic facet, and only parts of them align with the query. For instance, consider the query $(\texttt{David Beckham}, \texttt{has\_son}, ?)$, which seeks to identify family-related entities. In this case, embeddings that encode knowledge regarding ``family life'' (e.g., facts incorporating relations like \texttt{is\_married\_to} or \texttt{has\_sister}) evidently offer critical clues for the query. Conversely, embeddings emphasizing ``career trajectory'' (e.g., facts involving relations such as \texttt{played\_for} or \texttt{teammate}) or other domains provide less contribution or may even introduce noise by suggesting completely irrelevant entities.

With this in mind, we introduce a \textit{semantic-aware link prediction} method consisting of two steps: \textbf{(1) Semantic-aware importance computation}, which quantifies the semantic alignment between the embeddings $\textbf{E}_i$ of each snapshot and the current query $(h,r,?)$, and \textbf{(2) Link prediction with importance}, which aggregates the predictions from different snapshot embeddings using the computed importance to enhance the overall prediction accuracy.

\begin{table*}[!t]
\footnotesize
\centering
\renewcommand{\arraystretch}{1} 
\caption{Dataset statistics: $|E_{i}|$, $|R_{i}|$, and $|\Delta T_{i}|$ are the number of entities, relations, and new facts in each snapshot $S_i$}
\vspace{-0.25cm}
\resizebox{\textwidth}{!}{
\begin{tabular}{cccccccccccccccccccc}
\toprule
\multirow{2}{*}{\textbf{Datasets}} & \multicolumn{3}{c}{\textbf{Snapshot 1}} & & \multicolumn{3}{c}{\textbf{Snapshot 2}} & & \multicolumn{3}{c}{\textbf{Snapshot 3}} & & \multicolumn{3}{c}{\textbf{Snapshot 4}} & & \multicolumn{3}{c}{\textbf{Snapshot 5}} \\ 
\cline{2-4} \cline{6-8} \cline{10-12} \cline{14-16} \cline{18-20} 
 & \multicolumn{1}{c}{$|E_{1}|$} & \multicolumn{1}{c}{$|R_{1}|$} & \multicolumn{1}{c}{$|\Delta T_{1}|$} & & \multicolumn{1}{c}{$|E_{2}|$} & \multicolumn{1}{c}{$|R_{2}|$} & \multicolumn{1}{c}{$|\Delta T_{2}|$} & & \multicolumn{1}{c}{$|E_{3}|$} & \multicolumn{1}{c}{$|R_{3}|$} & \multicolumn{1}{c}{$|\Delta T_{3}|$} & & \multicolumn{1}{c}{$|E_{4}|$} & \multicolumn{1}{c}{$|R_{4}|$} & \multicolumn{1}{c}{$|\Delta T_{4}|$} & & \multicolumn{1}{c}{$|E_{5}|$} & \multicolumn{1}{c}{$|R_{5}|$} & \multicolumn{1}{c}{$|\Delta T_{5}|$} \\ \hline
ENTITY   & 2,909 & 223 & 46,388 && 5,817 & 236 & 72,111 && 8,275 & 236 & 73,785 && 11,633 & 237 & 70,506 && 14,541 & 237 & 47,326\\
RELATION & 11,560 & 48 & 98,819 && 13,343 & 96 & 93,535 && 13,754 & 143 & 66,136 && 14,387 & 190 & 30,032 && 14,541 & 237 & 21,594\\
HYBRID   & 8,628 & 86 & 57,561 && 10,040 & 102 & 20,873 && 12,779 & 151 & 88,017 && 14,393 & 209 & 103,339 && 14,541 & 237 & 40,326\\
FB-CKGE  & 7,505 & 237 & 186,070 && 11,258 & 237 & 31,012 && 13,134 & 237 & 31,012 && 14,072 & 237 & 31,012 && 14,541 & 237 & 31,010\\
GraphEqual & 2,908 & 226 & 57,636 && 5,816 & 235 & 62,023 && 8,724 & 237 & 62,023 && 11,632 & 237 & 62,023 && 14,541 & 237 & 66,411\\
GraphHigher & 900 & 197 & 10,000 && 1,838 & 221 & 20,000 && 3,714 & 234 & 40,000 && 7,467 & 237 & 80,000 && 14,541 & 237 & 160,116\\
GraphLower & 7,505 & 237 & 160,000 && 11,258 & 237 & 80,000 && 13,134 & 237 & 40,000 && 14,072 & 237 & 20,000 && 14,541 & 237 & 10,116\\
DB-CKGE  & 49,453 & 174 & 118,518 && 64,296 & 350 & 118,518 && 71,077 & 526 & 118,518 && 74,915 & 704 & 118,518 && 75,672 & 882 & 118,518\\
\bottomrule
\vspace{-0.65cm}
\end{tabular}
}
\label{tab:data_statistics}
\end{table*}

\vspace{0.1cm}
\noindent \textbf{Semantic-aware importance computation.}
The primary challenge lies in quantifying the relevance between the embeddings $\textbf{E}_i$ of each snapshot and the query $(h,r,?)$. Drawing on the intuition that ``an entity's semantics are reflected by its associated relations'' \cite{niu2024knowledge_relation_reflect}, we utilize the relations appearing in the new facts \QScolor{$\Delta D_i$} of $S_i$ as a proxy for the semantic focus of $\textbf{E}_i$. The underlying rationale is straightforward: the more semantically similar the relations in \QScolor{$\Delta D_i$} are to the relation $r$, the more likely $\textbf{E}_i$ encodes knowledge relevant to the query $(h,r,?)$. Revisiting the example query with $r=\texttt{has\_son}$, relations such as \texttt{is\_married\_to} and \texttt{has\_sister} that are from the family domain exhibit strong similarity to $r$. Consequently, the entities informed by those relations are of higher importance. Conversely, relations from unrelated domains (e.g., \texttt{played\_for} or \texttt{has\_hobby}) show clear semantic mismatch with the query, leading to lower importance for their associated entities.

Given a query $(h,r,?)$, we first extract the relations associated with the head entity $h$ from the incremental \QScolor{training} facts \QScolor{$\Delta D_i$}, denoting them as the \textit{candidate relations} $R_h^i$. 
\QScolor{Since a head prediction query $(?,r,t)$ operates in the reverse direction, we analogously extract the relations associated with the tail entity $t$ to form its corresponding candidate set $R_t^i$.}
\QScolor{To ensure the strict independence of the evaluation sets, we derive these candidate relations exclusively from the training subset. This entirely blinds the importance computation to any validation or test triples, guaranteeing that the model relies solely on explicitly observed knowledge during inference.} 
We then compute the cosine similarity between each relation $r_j\in R_h^i$ and the query relation $r$, forming the similarity set ${\rm Sim}(R_h^i,r)$,
\QScolor{which is defined as follows:}
\begin{equation}
{\rm Sim}(R_h^i,r) = \{sim(\textbf{r}_j, \textbf{r}) | r_j \in R_h^i\}\ .
\end{equation}

Since a single snapshot often encompasses a mixture of relations, including generic or weakly relevant ones that act as noise, 
\QScolor{considering all of them would dilute the truly relevant high-scoring signals. Conversely, selecting only the highest association fails to adequately capture the varying degrees of relevance across different snapshots. }
Therefore, we employ a Top-$k$ filtering strategy to assess the importance of $\textbf{E}_i$ by averaging only the top-$k$ highest similarities, denoted by $\delta_h^i$:
\begin{equation}
\delta_h^i = \frac{1}{k} \sum {\rm Top}k({\rm Sim}(R_h^i,r))\ .
\end{equation}

This strategy ensures that the importance of $\textbf{E}_i$ is driven by the most relevant semantics encoded within it. 
\QScolor{We further explore the impact of the parameter $k$ in the parameter sensitivity analysis of our experiments.}
\QScolor{Notably, if the candidate relation set is empty (i.e., $R_h^i = \emptyset$), we directly set the aggregated similarity $\delta_h^i$ to $0$.}
Finally, we employ the \textsf{Softmax} function to normalize the importance, denoted by $\beta_{h}^i$, which represents the contribution of $\textbf{E}_i$ to the given query:
\begin{equation}
\beta_{h}^i = \frac{exp(\delta_h^i)}{\sum_{j=1}^{n} exp(\delta_h^j)}\ .
\end{equation}

\noindent \textbf{Link prediction with importance.} Given a query $(h,r,?)$ and a growing KG $G=\{S_1,\dots,S_{n}\}$, we perform link prediction with importance as follows. For each visible entity $e\in E_i$ in a snapshot $S_i$, we form a candidate triple $(h,r,e)$ and compute the prediction score as $\psi^i_{(h,r,e)}$, where $p=1$ or $2$ represents the $L_1$ or $L_2$ norm. 
\QScolor{Notably, if the head $h$ or candidate entity $e$ lacks an explicit embedding in the current snapshot $E_i$ due to our Semantic Decoupling strategy for redundancy consolidation, we employ a pointer redirection strategy. Specifically, we retrieve the most similar historical embedding (e.g., $\textbf{e}^*_{x,j}$) via the pointer and directly substitute it into the equation to compute $\psi^i_{(h,r,e)}$.}
\begin{equation}
\psi^i_{(h,r,e)} = - \lVert\textbf{h} + \textbf{r} - \textbf{e}\rVert_{p}
\end{equation}

Since an entity $e$ can appear in multiple snapshots, there may be multiple prediction scores for candidate triples involving this entity—one from each snapshot. In such cases, we aggregate these predictions by weighting them according to the importance of each snapshot embedding $\textbf{E}_i$ w.r.t. the query (Eq. \ref{eq:agg_pre}), and then derive the final prediction with the greatest aggregated prediction score.
\begin{equation}
\label{eq:agg_pre}
\psi_{(h,r,e)} = \sum_{i=1}^{n} \beta_{h}^i\cdot \psi_{(h,r,e)}^i
\end{equation}

\begin{example}
    As shown in Figure \ref{fig:framework} (right part), suppose the importance of snapshot embeddings $\{\textbf{E}_1,\dots,\textbf{E}_3\}$ w.r.t. the query relation $r=\texttt{has\_son}$ is $\{0.8,0.1,0.1\}$, indicating that the semantic focus of $\textbf{E}_1$ lies primarily in the family-domain, while $\textbf{E}_2$ and $\textbf{E}_3$ emphasize other domains (just like the illustration in Figure \ref{fig:example}). Consequently, the embeddings from the first snapshot significantly boost the prediction scores of family-related entities, contributing more to answering the query $(\texttt{David Beckham}, \texttt{has\_son}, ?)$ and finally predicting \texttt{Brooklyn Beckham} rather than other irrelevant entities. In contrast, the embeddings from the second snapshot substantially enhance the scores of football-career-related entities, thereby playing a stronger role in queries such as $(\texttt{David Beckham}, \texttt{teammate}, ?)$.
\end{example}

\vspace{-0.3cm}
\section{Experiments}
We conducted experiments to answer the following questions. All experiments were conducted on a Linux server with a 2.4 GHz CPU, 128 GB memory, and an NVIDIA RTX A6000 GPU. Our code and datasets are available at the anonymous GitHub repository \citep{code}.

\vspace{0.05cm}
\noindent\textbf{Q1:} How does the proposed MF-CKGE perform in terms of effectiveness on the link prediction task compared to baselines? (\textbf{\S \ref{sec:effect}})

\vspace{0.05cm}
\noindent\textbf{Q2:} How does MF-CKGE compare to baselines in terms of training time and GPU memory consumption? (\textbf{\S \ref{sec:overhead}})

\vspace{0.05cm}
\noindent\textbf{Q3:} How sensitive is our MF-CKGE to parameter settings? (\textbf{\S \ref{sec:parameter}})

\vspace{0.05cm}
\noindent\textbf{Q4:} What is the contribution of each component to MF-CKGE's overall effectiveness (ablation study)? (\textbf{\S \ref{sec:ablation}})

\vspace{0.05cm}
\noindent\textbf{Q5:} How does MF-CKGE suppress query-irrelevant noise to enhance prediction accuracy (case study)? (\textbf{\S \ref{sec:case}})

    \begin{table*}[!t]
    \vspace{-0.1cm}
    \centering
    \footnotesize
    \renewcommand{\arraystretch}{1.1}
    \caption{Main results on the first four datasets. The best and second best results are highlighted in
    \textcolor{red}{bold} and \textcolor{blue}{\underline{underlined}}}
    \vspace{-0.3cm}
    \resizebox{0.88\textwidth}{!}{
    \begin{tabular}{cccccccccccccccc}
    \toprule
    \multirow{2}{*}{\textbf{Methods}} & \multicolumn{3}{c}{\textbf{ENTITY}} & & \multicolumn{3}{c}{\textbf{RELATION}} & & \multicolumn{3}{c}{\textbf{HYBRID}} & &\multicolumn{3}{c}{F\textbf{B-CKGE}} \\
    \cline{2-4} \cline{6-8} \cline{10-12} \cline{14-16}
    & MRR & Hits@1 & Hits@10 && MRR & Hits@1 & Hits@10 && MRR & Hits@1 & Hits@10 && MRR & Hits@1 & Hits@10 \\ \hline
     Retrain
     & 0.264 & 0.160 & 0.460 && 0.246 & 0.151 & 0.429 && 0.251 & 0.155 & 0.437 && 0.24 & 0.142 & 0.432 \\
     \hline
    PNN
    \cite{rusu2016progressive-PNN}
    & 0.229 & 0.130 & 0.425 && 0.167 & 0.096 & 0.305 && 0.185 & 0.101 & 0.349 && 0.215 & 0.122 & 0.403 \\
     CWR
     \cite{lomonaco2017core50-CWR}
     & 0.088 & 0.028 & 0.202 && 0.021 & 0.010 & 0.043 && 0.037 & 0.015 & 0.077 && 0.075 & 0.011 & 0.192 \\
     FastKGE
     \cite{liu2024fast-FastKGE}
     & 0.239 & 0.146 & 0.427 && 0.185 & 0.107 & 0.359 && 0.211 & 0.128 & 0.382 && 0.223 & 0.131 & 0.405 \\
     \hline
     GEM
     \cite{lopez2017gradient-GEM}
     & 0.165 & 0.085 & 0.321 && 0.093 & 0.040 & 0.196 && 0.136 & 0.070 & 0.263 && 0.188 & 0.103 & 0.359 \\
     EMR
     \cite{wang2019sentence-EMR}
     & 0.171 & 0.090 & 0.330 && 0.111 & 0.052 & 0.225 && 0.141 & 0.073 & 0.267 && 0.180 & 0.097 & 0.346 \\
     DiCGRL
     \cite{kou2020disentangle-DiCGRL}
     & 0.107 & 0.057 & 0.211 && 0.133 & 0.079 & 0.241 && 0.149 & 0.083 & 0.277 && 0.149 & 0.091 & 0.261 \\
    \hline
     SI
     \cite{zenke2017continual-SI}
     & 0.154 & 0.072 & 0.311 && 0.113 & 0.055 & 0.224 && 0.111 & 0.049 & 0.229 && 0.187 & 0.102 & 0.359 \\
     EWC
     \cite{kirkpatrick2017overcoming-EWC}
     & 0.229 & 0.130 & 0.423 && 0.165 & 0.093 & 0.306 && 0.186 & 0.102 & 0.350 && 0.218 & 0.124 & 0.410 \\
     LKGE
     \cite{cui2023lifelong-LKGE}
     & 0.234 & 0.136 & 0.425 && 0.192 & 0.106 & 0.366 && 0.207 & 0.121 & 0.379 && 0.208 & 0.113 & 0.403 \\
     IncDE
     \cite{liu2024towards-IncDE}
     & \textcolor{blue}{\underline{0.253}} 
    & \textcolor{blue}{\underline{0.151}} 
    & \textcolor{blue}{\underline{0.448}} 
    && \textcolor{blue}{\underline{0.199}} 
    & \textcolor{blue}{\underline{0.111}} 
    & \textcolor{blue}{\underline{0.370}} 
    && \textcolor{blue}{\underline{0.224}} 
    & \textcolor{blue}{\underline{0.131}} 
    & \textcolor{blue}{\underline{0.401}} 
    && \textcolor{blue}{\underline{0.232}} 
    & \textcolor{blue}{\underline{0.135}} 
    & \textcolor{blue}{\underline{0.423}} \\ \hline
    \textbf{MF-CKGE (Ours)} & \textbf{\textcolor{red}{0.261}} 
    & \textbf{\textcolor{red}{0.160}} 
    & \textbf{\textcolor{red}{0.452}} 
    && \textbf{\textcolor{red}{0.226}} 
    & \textbf{\textcolor{red}{0.141}} 
    & \textbf{\textcolor{red}{0.389}} 
    && \textbf{\textcolor{red}{0.244}} 
    & \textbf{\textcolor{red}{0.151}} 
    & \textbf{\textcolor{red}{0.418}} 
    && \textbf{\textcolor{red}{0.241}} 
    & \textbf{\textcolor{red}{0.146}} 
    & \textbf{\textcolor{red}{0.423}} \\
    \bottomrule
    \end{tabular}
    }
    \label{tab:result1}
    \vspace{-0.2cm}
    \end{table*}

\begin{table*}[!t]
\centering
\footnotesize
\renewcommand{\arraystretch}{1.1}
\caption{Main results on the last four datasets. The best and second best results are highlighted in
\textcolor{red}{bold} and \textcolor{blue}{\underline{underlined}}}
\vspace{-0.3cm}
\resizebox{0.88\textwidth}{!}{
\begin{tabular}{cccccccccccccccc}
\toprule
\multirow{2}{*}{\textbf{Methods}} & \multicolumn{3}{c}{\textbf{GraphEqual}} && \multicolumn{3}{c}{\textbf{GraphHigher}} && \multicolumn{3}{c}{\textbf{GraphLower}} && \multicolumn{3}{c}{\textbf{DB-CKGE}} \\
\cline{2-4} \cline{6-8} \cline{10-12} \cline{14-16} 
 & MRR    & Hits@1    & Hits@10  && MRR    & Hits@1    & Hits@10  && MRR    & Hits@1   & Hits@10  && MRR   & Hits@1   & Hits@10 \\ \hline
 Retrain 
 & 0.242 & 0.14 & 0.444 && 0.252 & 0.148 & 0.454 && 0.23 & 0.132 & 0.425 && 0.19 & 0.082 & 0.376 \\ \hline
PNN 
\cite{rusu2016progressive-PNN}  
& 0.212  & 0.118 & 0.405 && 0.153  & 0.055 & 0.333 && 0.213  & 0.119 & 0.407 && 0.062 & 0.024 & 0.128 \\
CWR 
\cite{lomonaco2017core50-CWR}  
& 0.122  & 0.041  & 0.277 && 0.203  & 0.106  & 0.388 & & 0.032  & 0.005 & 0.080 && 0.026 & 0.006 & 0.059 \\
FastKGE 
\cite{liu2024fast-FastKGE}
& 0.213 & 0.124 & 0.387 && 0.187 & 0.109 & 0.339 && 0.215 & 0.124 & 0.396 && 0.043 & 0.014 & 0.095 \\
\hline
GEM 
\cite{lopez2017gradient-GEM} 
& 0.189  & 0.099  & 0.372 && 0.197  & 0.109  & 0.372 && 0.170  & 0.084 & 0.346 && 0.071 & 0.019 & 0.163 \\
EMR 
\cite{wang2019sentence-EMR}  
& 0.185  & 0.099  & 0.359 && 0.202  & 0.113  & 0.379 && 0.188  & 0.101 & 0.362 && 0.083 & 0.026 & 0.185 \\
DiCGRL
\cite{kou2020disentangle-DiCGRL} 
& 0.104  & 0.040  & 0.226 && 0.116  & 0.041  & 0.242 && 0.102  & 0.039 & 0.222 && 0.048 & 0.011 & 0.112 \\ \hline
SI 
\cite{zenke2017continual-SI}   
& 0.179  & 0.092  & 0.353 && 0.190  & 0.099 & 0.371 && 0.186  & 0.099 & 0.366 && 0.075 & 0.02 & 0.172 \\
EWC 
\cite{kirkpatrick2017overcoming-EWC} 
& 0.207  & 0.113  & 0.400 && 0.198  & 0.106  & 0.385 && 0.210  & 0.116 & 0.405 && 0.088 & 0.023 & 0.210 \\
LKGE 
\cite{cui2023lifelong-LKGE} 
& 0.214  & 0.118  & 0.407 && 0.207  & 0.120  & 0.382 && 0.210  & 0.116 & 0.403 && \textcolor{blue}{\underline{0.119}} & \textcolor{red}{\bf{0.053}} & \textcolor{blue}{\underline{0.236}} \\

IncDE 
\cite{liu2024towards-IncDE} 
& \textcolor{blue}{\underline{0.234}}  & \textcolor{blue}{\underline{0.134}} & \textcolor{blue}{\underline{0.432}} && \textcolor{blue}{\underline{0.227}}  & \textcolor{blue}{\underline{0.132}}  & \textcolor{blue}{\underline{0.412}} && \textcolor{blue}{\underline{0.228}} & \textcolor{blue}{\underline{0.129}} & \textcolor{red}{\bf{0.426}} && 0.081 & 0.040 & 0.157 \\
\hline

\textbf{MF-CKGE (Ours)}   & \textcolor{red}{\bf{0.247}}  & \textcolor{red}{\bf{0.146}}  & \textcolor{red}{\bf{0.441}} && \textcolor{red}{\bf{0.254}}  & \textcolor{red}{\bf{0.152}}  & \textcolor{red}{\bf{0.450}} && \textcolor{red}{\bf{0.233}}  & \textcolor{red}{\bf{0.136}} & \textcolor{blue}{\underline{0.423}} && \textcolor{red}{\bf{0.125}} & \textcolor{blue}{\underline{0.044}} & \textcolor{red}{\bf{0.266}} \\
\bottomrule
\end{tabular}
}
\vspace{-0.2cm}
\label{tab:result2}
\end{table*}

\subsection{Experimental Setup}
\noindent\textbf{Datasets.}
Consistent with prior work in CKGE literature, we employ seven widely used datasets, each featuring five snapshots and distinct evolutionary patterns. Table \ref{tab:data_statistics} summarizes the dataset statistics. In the ENTITY, RELATION, and HYBRID datasets \cite{cui2023lifelong-LKGE}, the number of entities, relations, and the mixture of both increase uniformly at each snapshot. Especially in RELATION and HYBRID, each incremental update introduces a substantial number of new facts from different domains. FB-CKGE \cite{liu2024fast-FastKGE} simulates real-world KG evolution by adding only a small portion of new facts to a foundational KG at each snapshot. In contrast, GraphEqual, GraphHigher, and GraphLower \cite{liu2024towards-IncDE} exhibit varying growth trends in the size of new facts: nearly constant in GraphEqual, increasing in GraphHigher, and decreasing in GraphLower. Despite these differences in incremental scale, all datasets show monotonic growth in cumulative triples. Furthermore, to assess MF-CKGE on larger and more semantically diverse KGs, we construct DB-CKGE, a growing KG derived from DBpedia \cite{auer2007dbpedia-DBpedia}. Compared to the above datasets, DB-CKGE has three distinguishing characteristics: (1) it has the largest KG at every snapshot, (2) the size of new facts at each snapshot remains at a consistently high level, and (3) the new facts cover a broader range of semantic domains. For all datasets, the training, validation, and test sets are allocated as 3:1:1 for each snapshot.

\vspace{0.05cm}
\noindent\textbf{Baselines.}
We compare MF-CKGE with ten representative baselines from three categories: (1) regularization-based, (2) dynamic architecture-based, and (3) replay-based methods. We also implement a fully retrained baseline \textsf{Retrain}. Although \textsf{Retrain} incurs prohibitive computational and memory overhead--making it impractical in real-world scenarios--\textbf{it provides a reference for expected performance under full-data access}, allowing us to assess how closely other methods approach this desirable yet costly regime.


\vspace{0.05cm}
\noindent\underline{(1) Regularization-based methods.} We compare four methods from this category: \textsf{IncDE} \cite{liu2024towards-IncDE}, \textsf{LKGE} \cite{cui2023lifelong-LKGE}, \textsf{SI} \cite{zenke2017continual-SI}, and \textsf{EWC} \cite{kirkpatrick2017overcoming-EWC}.

\noindent$\bullet$ \textbf{\textsf{IncDE}}. It adapts to the evolving nature of KGs through hierarchical processing and incremental distillation, ensuring that previous knowledge is effectively transferred to the current model.

\noindent$\bullet$ \textbf{\textsf{LKGE}}. It employs fact-frequency-based regularization to stabilize embeddings and mitigate forgetting, while utilizing a Masked Autoencoder and knowledge transfer to adapt to new knowledge.

\noindent$\bullet$ \textbf{\textsf{SI}}. It evaluates parameter importance by calculating the contribution to loss reduction, selectively restricting updates to these critical parameters to preserve prior knowledge.

\noindent$\bullet$ \textbf{\textsf{EWC}}. It uses regularization on the Fisher Information Matrix-based parameter importance updates to mitigate forgetting.

\vspace{0.05cm}
\noindent\underline{(2) Dynamic architecture-based methods.} We compare three representative methods: \textsf{FastKGE} \cite{liu2024fast-FastKGE}, \textsf{CWR} \cite{lomonaco2017core50-CWR}, and \textsf{PNN} \cite{rusu2016progressive-PNN}.

\noindent$\bullet$ \textbf{\textsf{FastKGE}}. It handles KG growth through an incremental low-rank adapter (IncLoRA), which dynamically allocates rank resources to embed new knowledge with additional parameter overhead.

\noindent$\bullet$ \textbf{\textsf{CWR}}. It mitigates forgetting using a dual-weight mechanism to isolate new learning in temporary weights before consolidating them to preserve class-specific knowledge.

\noindent$\bullet$ \textbf{\textsf{PNN}}. It instantiates a new network for each task, transferring knowledge from the frozen features of previous networks.

\vspace{0.05cm}
\noindent\underline{(3) Replay-based methods.} We compare three representative methods from this area: DiCGRL \cite{kou2020disentangle-DiCGRL}, EMR \cite{wang2019sentence-EMR}, and GEM \cite{lopez2017gradient-GEM}.

\noindent$\bullet$ \textbf{\textsf{DiCGRL}}. It disentangles node embeddings into independent components and prevents forgetting by dynamically replaying partial historical data to update only the relevant components.

\noindent$\bullet$ \textbf{\textsf{EMR}}. It preserves historical knowledge by augmenting the current training data with randomly sampled previously trained data.

\noindent$\bullet$ \textbf{\textsf{GEM}}. It stores past training data to detect knowledge conflicts and rotates the gradient to avoid forgetting the old knowledge.

\vspace{0.05cm}
\noindent\textbf{Queries.} We design experiments based on the lifelong link prediction task mentioned in \S \ref{sec:preliminary}. For each snapshot $S_i$ ($i\in [1,n]$) of a dataset, we divide it into a training set ($D_i$), a validation set ($V_i$), and a test set ($Q_i$) in a ratio of 3:1:1. We next form prediction queries on the accumulated test sets $\cup^n_{j=1}Q_j$. Specifically, for each test fact $(h, r, t)$ in $\cup^n_{j=1}Q_j$, we generate two types of queries: $(h, r, ?)$ to predict the tail entity, and $(?, r, t)$ to predict the head entity.

\begin{figure*}
    \vspace{-0.2cm}
    \centering
    \includegraphics[width=1\textwidth]{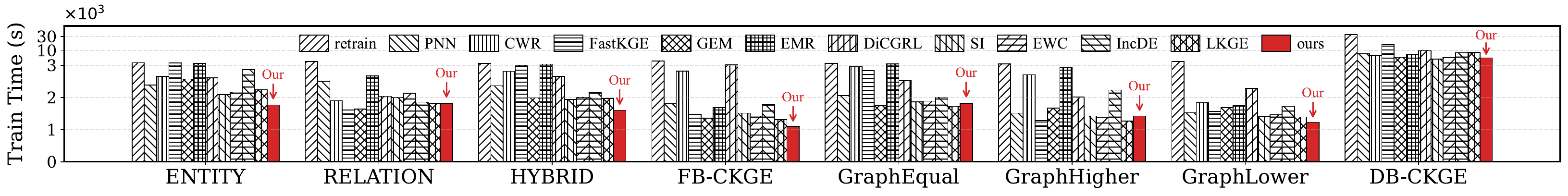}
    \vspace{-0.9cm}
    \caption{Training time of all methods on eight datasets}
    \label{fig:train time}
    \vspace{-0.4cm}
\end{figure*}

\begin{figure*}
    \centering
    \includegraphics[width=1\textwidth]{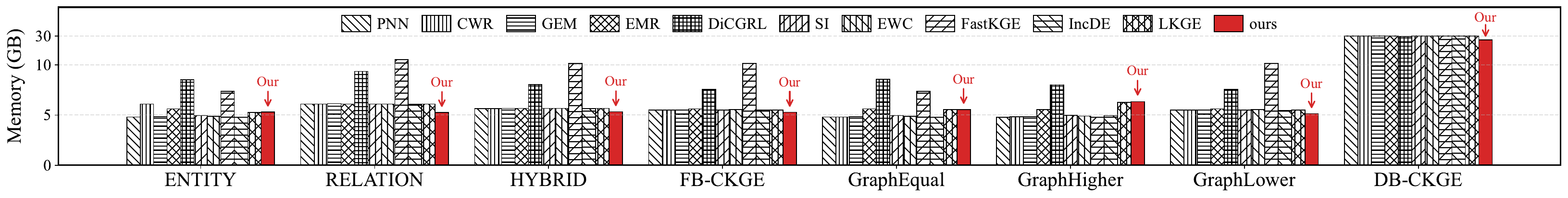}
    \vspace{-0.9cm}
    \caption{GPU memory usage of all methods on eight datasets}
    \label{fig:train memory}
    \vspace{-0.4cm}
\end{figure*}

\vspace{0.05cm}
\noindent\textbf{Metrics.} \QScolor{Following standard evaluation protocols in CKGE \cite{cui2023lifelong-LKGE,liu2024towards-IncDE}, we evaluate the effectiveness using three typical metrics: Mean Reciprocal Rank (MRR), Hits@1, and Hits@10 under filtered setting.}
\QScolor{Specifically, during the evaluation at $S_i$, for a test query $(h,r,?)$ (or $(?,r,t)$), the candidate set for ranking consists of all entities observed up to $S_i$. Furthermore, before ranking these candidates, we filter out all other true triples that appear in the accumulated training, validation, and test sets up to $S_i$, ensuring that other genuine facts do not inappropriately penalize the rank of the target entity.}
Higher values indicate better link prediction performance. The main results are averaged over five runs of the model trained on the last snapshot on queries. Moreover, we evaluate the training overhead in terms of time consumption and GPU memory usage.

\begin{figure}[t]
    \centering
    \includegraphics[width=0.94\linewidth]{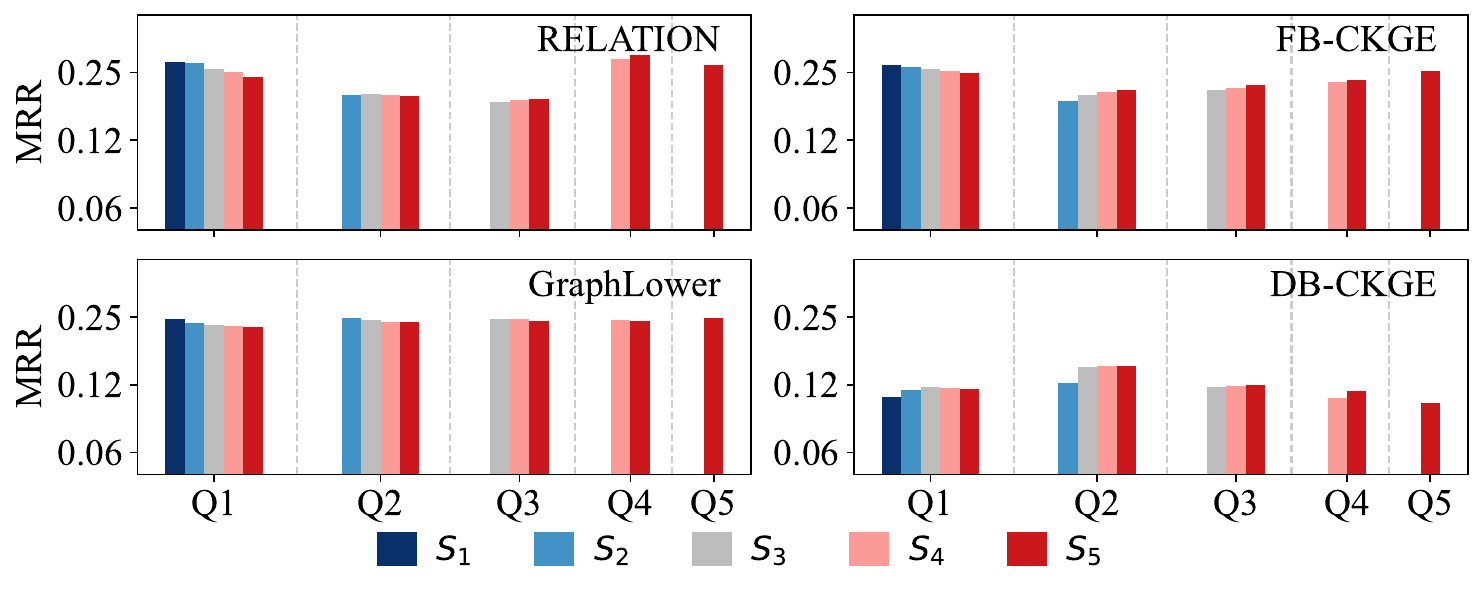}
    \vspace{-0.35cm}
    \caption{Effectiveness of MF-CKGE for retaining old knowledge on RELATION, FB-CKGE, GraphLower and DB-CKGE} 
    \vspace{-0.3cm}
    \label{fig:MRRsin snapshots}
\end{figure}

\begin{figure}
    \centering
    \includegraphics[width=1\linewidth]{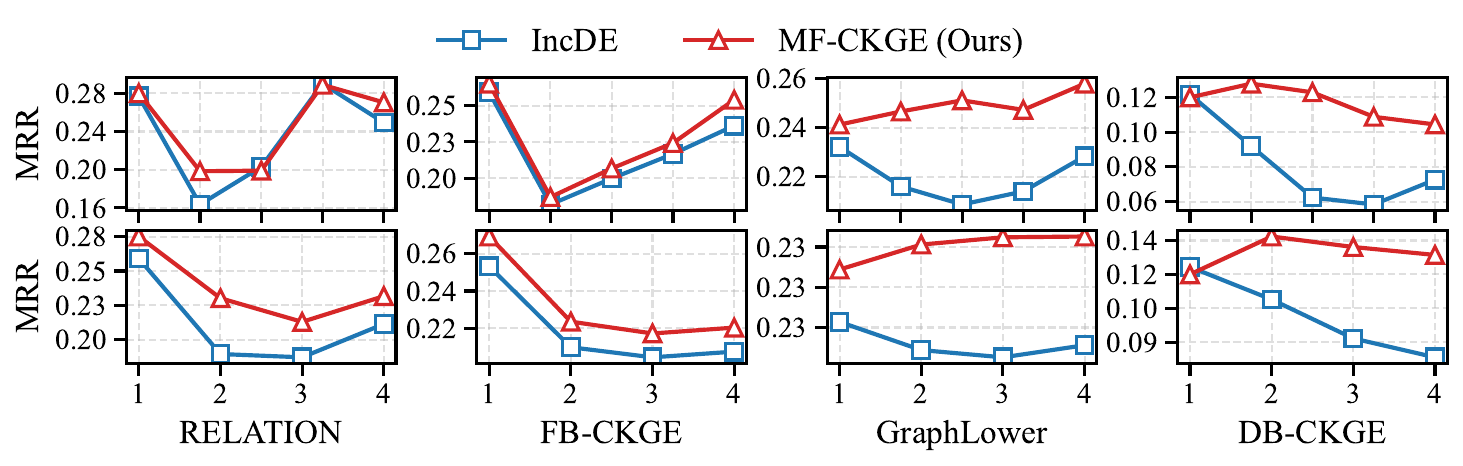}
    \vspace{-0.6cm}
    \caption{Effectiveness of learning new knowledge (top) and memorizing old knowledge (bottom)}
    \vspace{-0.4cm}
    \label{fig:mrr_our_vs_LKGE}
\end{figure}

\vspace{0.05cm}
\noindent\textbf{Settings.} The learning rate is chosen from \{0.0001, 0.0005, 0.001\}, and the batch size from \{512, 1024, 2048\}. We employ the Adam optimizer throughout. To ensure fair comparison with baselines, we adopt TransE \cite{bordes2013translating-KGE1} as the default base KGE model. The hyperparameters $\alpha$ and $\eta$ are selected from \{0.01, 0.1, 1.0\}, and the embedding dimensions for entities and relations are fixed at 200.

\subsection{Effectiveness Analysis}
\label{sec:effect}
\noindent\textbf{Main results.} We report the main results over eight datasets in Tables \ref{tab:result1} and \ref{tab:result2}.
Notably, \textbf{MF-CKGE outperforms all competing baselines, achieving an average (maximum) improvement of 1.7\% (2.7\%) in MRR, 1.3\% (3.0\%) in Hits@1, and 1.4\% (3.8\%) in Hits@10 over the best baseline}, demonstrating its general effectiveness. In detail, compared to dynamic architecture-based methods (\textsf{PNN}, \textsf{CWR}, and \textsf{FastKGE}), ours achieves improvements of 1.8\%$\sim$20.7\% in MRR, 1.2\%$\sim$13.6\% in Hits@1, and 1.6\%$\sim$34.6\% in Hits@10. Against replay-based methods (\textsf{GEM}, \textsf{EMR}, and \textsf{DiCGRL}), ours gains 4.3\%$\sim$15.4\% in MRR, 3.5\%$\sim$11.1\% in Hits@1, and 6.1\%$\sim$24.1\% in Hits@10. While for regularization-based methods (\textsf{SI}, \textsf{EWC}, \textsf{LKGE}, and \textsf{IncDE}), ours gains 0.8\%$\sim$13.3\%, 0.7\%$\sim$10.2\%, and 0.4\%$\sim$18.9\% in MRR, Hits@1, and Hits@10, respectively. 
Remarkably, MF-CKGE even surpasses \textsf{Retrain} on certain metrics over FB-CKGE and GraphHigher. This is attributed to its ability to capture multi-faceted entity semantics and amplify query-relevant entity embeddings through a semantic-aware importance weighting. Although \textsf{Retrain} typically achieves the best results, it incurs significantly higher training cost compared to all other methods. For instance, on FB-CKGE, \textsf{Retrain} takes 899s for training $S_5$, \textbf{whereas ours completes in just 46 seconds (19$\times$ faster) while maintaining competitive accuracy (MRR: 0.1\% higher, Hits@1: 0.4\% higher)}. More results of training overhead are shown in \S \ref{sec:overhead}.

\vspace{0.1cm}
\noindent\textbf{Retention of old knowledge.} We evaluate MF-CKGE trained on $S_i$ using accumulated test data $\cup_{j=1}^{i-1}Q_j$ to assess its ability to retain old knowledge. Each $Q_j$ is evaluated across $n-j$ subsequent snapshots ($n$ is the total number of snapshots). For example, $Q_1$ is tested on $S_2\sim S_5$, and the results reflect how well knowledge w.r.t. $Q_1$ is retained over time. As shown in Figure \ref{fig:MRRsin snapshots}, the results of each \( Q_j \) remain stable across different \( S_i \). Notably, on some datasets, evaluating earlier \( Q_j \) on more recent snapshots even improves results. 
For instance, $Q_2$ in DB-CKGE (from $S_3$ to $S_5$) and $Q_2 \sim Q_4$ in FB-CKGE exhibit consistent upward trends.
This suggests that MF-CKGE not only effectively retains old knowledge but also leverages new information to enhance predictions on old test sets.

\begin{table*}[!t]
\centering
\footnotesize
\caption{Ablation results on ENTITY, RELATION, HYBRID, and FB-CKGE}
\vspace{-0.2cm}
\renewcommand{\arraystretch}{1.1}
\resizebox{0.88\textwidth}{!}{
\begin{tabular}{cccccccccccccccc}
\toprule
\multirow{2}{*}{\textbf{Methods}} & \multicolumn{3}{c}{\textbf{ENTITY}} && \multicolumn{3}{c}{\textbf{RELATION}} && \multicolumn{3}{c}{\textbf{HYBRID}} && \multicolumn{3}{c}{\textbf{FB-CKGE}} \\ 
\cline{2-4} \cline{6-8} \cline{10-12} \cline{14-16}
 & MRR & Hits@1 & Hits@10 && MRR & Hits@1 & Hits@10 && MRR & Hits@1 & Hits@10 && MRR & Hits@1 & Hits@10 \\ \hline 
MF-CKGE & 0.261 & 0.160 & 0.452 && 0.226 & 0.141 & 0.389 && 0.244 & 0.151 & 0.418 && 0.241 & 0.146 & 0.423 \\ 
w/o decoupling
& 0.259 & 0.158 & 0.451 && 0.211  & 0.122  & 0.383 && 0.239  & 0.145 & 0.416 && 0.238 & 0.135 & 0.422 \\
w/o importance
& 0.240 & 0.142 & 0.423 && 0.203  & 0.119  & 0.369 && 0.230  & 0.139 & 0.403 && 0.222 & 0.131 & 0.398 \\
\bottomrule
\end{tabular} 
}
\vspace{-0.1cm}
\label{tab:ablation1}
\end{table*}

\begin{table*}[!t]
\centering
\footnotesize
\caption{Ablation results on GraphEqual, GraphHigher, GraphLower, and DB-CKGE} 
\vspace{-0.3cm}
\renewcommand{\arraystretch}{1.1}
\resizebox{0.88\textwidth}{!}{
\begin{tabular}{cccccccccccccccc}
\toprule
\multirow{2}{*}{\textbf{Methods}} & \multicolumn{3}{c}{\textbf{GraphEqual}} && \multicolumn{3}{c}{\textbf{GraphHigher}} && \multicolumn{3}{c}{\textbf{GraphLower}} && \multicolumn{3}{c}{\textbf{DB-CKGE}} \\ 
\cline{2-4} \cline{6-8} \cline{10-12} \cline{14-16} 
 & MRR & Hits@1 & Hits@10 && MRR & Hits@1 & Hits@10 && MRR & Hits@1 & Hits@10 && MRR & Hits@1 & Hits@10 \\ \hline 
MF-CKGE & 0.247 & 0.146 & 0.441 
&& 0.254 & 0.152 & 0.450
&& 0.233 & 0.136 & 0.423
&& 0.125 & 0.044 & 0.266 \\ 
w/o decoupling
& 0.246  & 0.143  & 0.441
&& 0.239  & 0.141  & 0.432 
&& 0.232  & 0.134 & 0.421 
&& 0.099  & 0.038  & 0.212 \\
w/o importance
& 0.232  & 0.134  & 0.426 
&& 0.248  & 0.148  & 0.440 
&& 0.214  & 0.121 & 0.397 
&& 0.108  & 0.037  & 0.234 \\
\bottomrule
\end{tabular} 
}
\vspace{-0.3cm}
\label{tab:ablation2} 
\end{table*}

\vspace{0.1cm}
\noindent\textbf{Effect of Learning and Memorizing.} 
To verify that MF-CKGE can learn new knowledge well and retain old knowledge, we compare its performance with the best baseline, \textsf{IncDE}, as shown in Figure~\ref{fig:mrr_our_vs_LKGE}. Regarding the learning of new knowledge (top row), MF-CKGE consistently outperforms \textsf{IncDE} across most snapshots. Notably, on the DB-CKGE dataset, MF-CKGE demonstrates a significant advantage, maintaining a high MRR while \textsf{IncDE} shows a sharp decline. This indicates that our decoupling strategy effectively allows the model to capture new knowledge accurately. In terms of retaining old knowledge (bottom row), MF-CKGE achieves consistently higher MRR scores than \textsf{IncDE}. Both demonstrate that MF-CKGE effectively mitigates the issue of catastrophic forgetting.

\vspace{-0.2cm}
\subsection{Training Overhead}
\label{sec:overhead}
\noindent\textbf{Training time.} Figure \ref{fig:train time} shows the training time of all methods across eight datasets.
Our MF-CKGE incurs the lowest time cost, \textbf{achieving an 8$\times$ speedup compared to \textsf{Retrain}}, which requires accessing the entire KG. Moreover, \textbf{compared with the strongest baseline \textsf{IncDE}, ours requires less training time (a 1.4$\times$ speedup)}. This gain stems primarily from our design to focus training exclusively on new knowledge, avoiding the computational overhead arising from the complex trade-off between historical preservation and new knowledge acquisition.

\begin{figure}[t]
    \centering
    \includegraphics[width=1\linewidth]{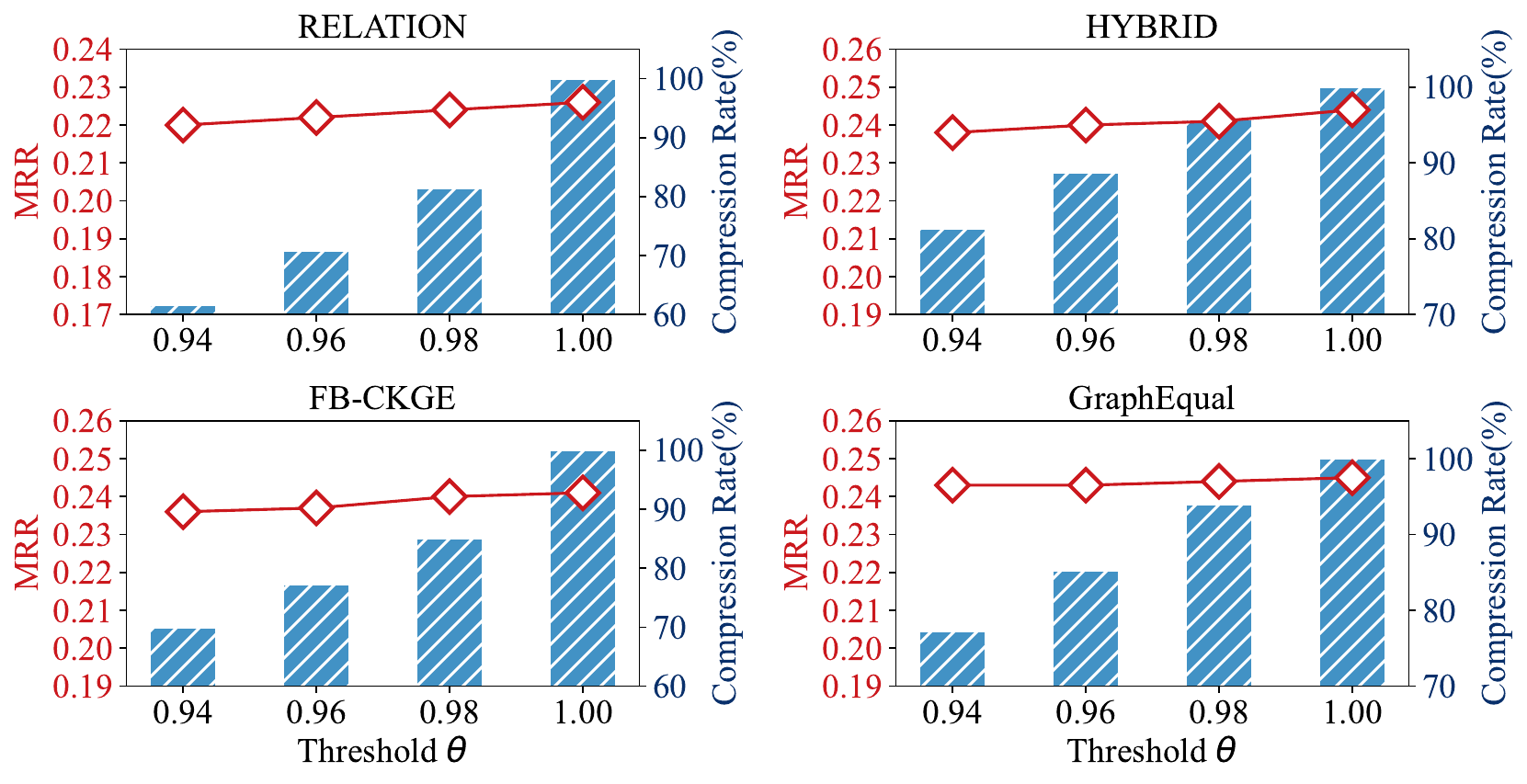}
    \vspace{-0.6cm}
    \caption{Effect of $\theta$ on space compression and effectiveness}
    \vspace{-0.6cm}
    \label{fig:compress}
\end{figure}

\vspace{0.1cm}
\noindent\textbf{GPU memory consumption.} We also present the runtime GPU memory usage of all methods across eight datasets in Figure \ref{fig:train memory}. Except for \textsf{DiCGRL} and \textsf{FastKGE}, which exhibit higher memory demands than others, our MF-CKGE maintains a memory footprint comparable to most baselines, without introducing significant additional overhead. This is primarily because most methods require allocating embedding spaces for all entities and relations, as well as processing training samples with a fixed batch size.

\begin{table*}[t]
\centering 
\footnotesize
\caption{Effect of various KGE models over ENTITY, RELATION, HYBRID, and FB-CKGE} 
\vspace{-0.3cm}
\renewcommand{\arraystretch}{1.1}
\resizebox{0.88\textwidth}{!}{
\begin{tabular}{cccccccccccccccccccc}
\toprule
\multirow{2}{*}{\textbf{Methods}}& \multicolumn{4}{c}{\textbf{ENTITY}} && \multicolumn{4}{c}{\textbf{RELATION}} && \multicolumn{4}{c}{\textbf{HYBRID}}&& \multicolumn{4}{c}{\textbf{FB-CKGE}}\\ 
\cline{2-5} \cline{7-10} \cline{12-15} \cline{17-20}
 & MRR & Hits@1 & Hits@10 & Time (s) && MRR & Hits@1 & Hits@10 & Time (s) && MRR & Hits@1 & Hits@10 & Time (s)&& MRR & Hits@1 & Hits@10 & Time (s)\\ \hline 

MF-CKGE w/ TransE
& 0.261 & 0.160 &  0.452 & 1,765

&& 0.226  & 0.141 & 0.389 & 1,816 
&& 0.244  & 0.151& 0.418 & 1,602
&& 0.241  & 0.146 & 0.423 & 1,103\\ 
MF-CKGE w/ RotatE
& 0.372 & 0.256 & 0.595 & 4,022

&& 0.281 & 0.176  & 0.490 & 4,236
&& 0.320 & 0.209  & 0.536 & 5,553
&& 0.309 & 0.204  & 0.514 & 3,019\\
MF-CKGE w/ ComplEx
& 0.317 & 0.218 & 0.519 & 4,676

&& 0.306  & 0.215 & 0.491 & 3,368
&& 0.278  & 0.181 & 0.478 & 6,530
&& 0.321  & 0.216 & 0.536 & 1,147\\
\bottomrule
\end{tabular} 
}
\label{tab:baseKGE1}
\vspace{-6pt}
\end{table*}

\begin{table*}[t]
\centering 
\footnotesize
\caption{Effect of various KGE models over GraphEqual, GraphHigher, GraphLower, and DB-CKGE} 
\vspace{-0.3cm}
\renewcommand{\arraystretch}{1.1}
\resizebox{0.88\textwidth}{!}{
\begin{tabular}{cccccccccccccccccccc}
\toprule
\multirow{2}{*}{\textbf{Methods}}& \multicolumn{4}{c}{\textbf{GraphEqual}} && \multicolumn{4}{c}{\textbf{GraphHigher}} && \multicolumn{4}{c}{\textbf{GraphLower}}&& \multicolumn{4}{c}{\textbf{DB-CKGE}}\\ 
\cline{2-5} \cline{7-10} \cline{12-15} \cline{17-20}
 & MRR & Hits@1 & Hits@10 & Time (s) && MRR & Hits@1 & Hits@10 & Time (s) && MRR & Hits@1 & Hits@10 & Time (s)&& MRR & Hits@1 & Hits@10 & Time (s)\\ \hline 

MF-CKGE w/ TransE
& 0.247 & 0.146&  0.441 & 1,822

&& 0.254  & 0.152& 0.450 & 1,422 
&& 0.233  & 0.136& 0.423 & 1,227
&& 0.125  & 0.044& 0.266 & 5,464\\ 
MF-CKGE w/ RotatE
& 0.322 & 0.204 & 0.561 & 2,846

&& 0.339  & 0.219 & 0.582 & 3,211
&& 0.287 & 0.185  & 0.497 & 2,976
&& 0.179 & 0.093  & 0.341 & 16,660\\
MF-CKGE w/ ComplEx
& 0.321  & 0.177 & 0.536 & 1,945

&& 0.279  & 0.188 & 0.468 & 2,334
&& 0.299  & 0.197 & 0.514 & 2,545
&& 0.205  & 0.136 & 0.337 & 8,759\\
\bottomrule
\end{tabular} 
}
\label{tab:baseKGE2} 
\vspace{-6pt}
\end{table*}

\vspace{-0.2cm}
\subsection{Parameter Sensitivity}
\label{sec:parameter}
\noindent\textbf{Effect of the threshold $\theta$ in semantic decoupling.} Figure \ref{fig:compress} shows the effect of $\theta$ on the model size (compression ratio, right $Y$-axis) and effectiveness (MRR, left $Y$-axis) across RELATION, HYBRID, FB-CKGE and GraphEqual. The compression ratio is the post-compression model size relative to the original size. As $\theta$ decreases, more space is reduced, resulting in a smaller compression ratio. Despite this notable reduction in model space, MRR remains largely stable. 
\QScolor{Even under the aggressive setting of $\theta=0.94$ on FB-CKGE, where about 30\% of the space is removed, MF-CKGE maintains strong performance, with the MRR decreasing slightly from 24.1\% to 23.6\%, which still outperforms \textsf{IncDE}.}
This confirms that semantic decoupling can effectively reduce model size without sacrificing effectiveness, demonstrating its robustness.

\begin{figure}[t]
    \centering
    \includegraphics[width=1\linewidth]{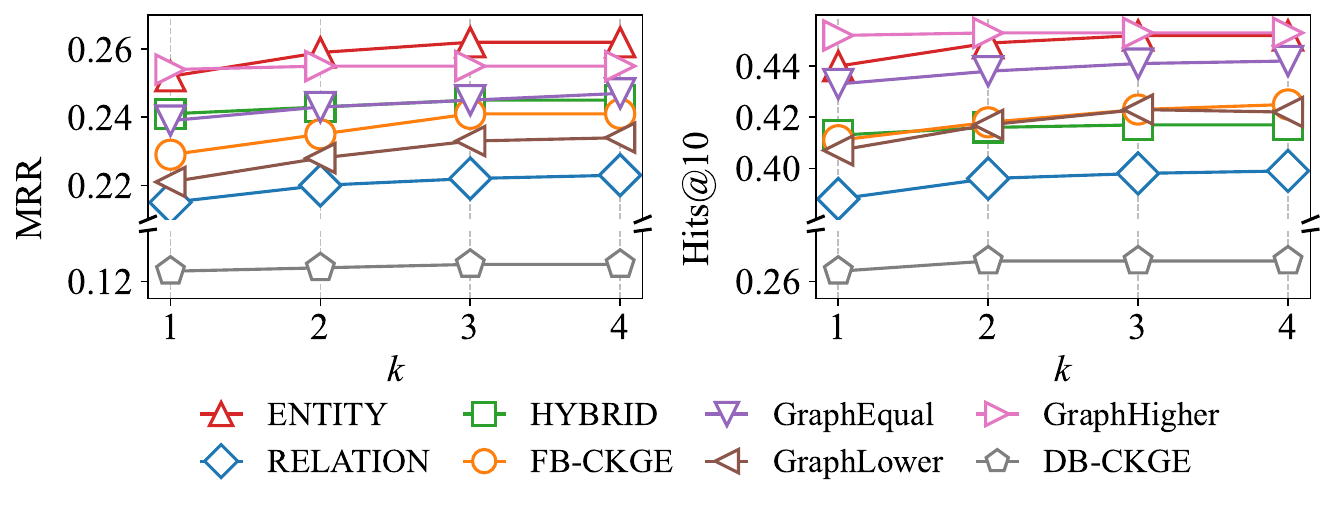}
    \vspace{-0.8cm}
    \caption{Effect of $k$ on effectiveness}
    \label{fig:top-k}
    \vspace{-0.6cm}
\end{figure}

\vspace{0.1cm}
\noindent\textbf{Effect of Top-$k$ selection in importance computation.}
Figure \ref{fig:top-k} shows that as $k$ increases from 1 to 4, the MRR and Hits@10 consistently improve. However, at $k=4$, the performance generally plateaus across most datasets, and in some cases even declines. For example, in FB-CKGE, MRR rises from 0.23 at $k=1$ to 0.24 at $k=3$, while remaining stable at $k=4$; Hits@10 on GraphLower even exhibits a decrease at $k=4$. These results underscore the importance of choosing an appropriate $k$ (e.g., $k=3$): averaging the top-$k$ similarities effectively aggregates strong semantic clues and filters out coincidental high-similarity noise; however, an overly large $k$ can dilute semantic perception and degrade prediction performance.

\vspace{-0.4cm}
\subsection{Ablation Study}
\label{sec:ablation}
\noindent\textbf{Ablation results.}
Table \ref{tab:ablation1} and Table \ref{tab:ablation2} examine the effect of two modules, \textit{entity embeddings decoupling} (M1) and \textit{semantic-aware importance computation} (M2). Removing M1 (i.e., new knowledge is no longer separated from old one) degrades performance across most datasets.
\QScolor{The most significant declines occur on DB-CKGE, RELATION, and GraphHigher, where the MRR drops by 2.6\% (from 12.5\% to 9.9\%), 1.5\% (from 22.6\% to 21.1\%), and 1.5\% (from 25.4\% to 23.9\%), respectively.}
This indicates that decoupling facilitates more effective encoding of both new and old knowledge, preventing semantic entanglement, especially in scenarios with significant updates. Removing M2 reduces performance across all datasets. 
\QScolor{Notably, without the semantic-aware importance weighting, the MRR and Hits@10 on ENTITY experience substantial drops of 2.1\% (26.1\% to 24.0\%) and 2.9\% (45.2\% to 42.3\%), respectively. A similar trend is observed on RELATION, with MRR dropping by 2.3\% (22.6\% to 20.3\%).}
This confirms its critical role in dynamically guiding the model to focus on the most query-relevant embeddings and suppressing semantic noise during prediction.

\vspace{0.1cm}
\noindent\textbf{Effect of various KGE models.} Since most baselines are built upon TransE and incompatible with other KGE models, we adopt TransE as the default for fair comparison. In contrast, MF-CKGE is extensible to other KGE models. So, we study the effect of KGE models using the bilinear-based ComplEx \cite{trouillon2016complex-Complex} and rotation-translation-based RotatE \cite{sun2019rotate-RotatE}. In Table \ref{tab:baseKGE1} and Table \ref{tab:baseKGE2}, both KGE models consistently outperform that using TransE (e.g., average +3.2\% in MRR). However, they also incur much higher training costs (e.g., average $2.4\times$ longer time), making the lightweight TransE a more practical choice that offers a better trade-off in general-purpose settings.

\begin{table}[t]
\centering
\caption{Case study on ENTITY}
\vspace{-0.3cm}
\renewcommand{\arraystretch}{1.1}
\resizebox{\columnwidth}{!}{
\begin{tabular}{llll} 
\toprule
\multicolumn{4}{c}{\textbf{Case 1:} Query (Harvard School of Law, field\_of\_study, ?)} \\
\hline
Semantic facet & \multicolumn{1}{c}{Academic Fields} & \multicolumn{1}{c}{School Symbols} & \multicolumn{1}{c}{Graduates} \\
\hline
Importance & \multicolumn{1}{c}{0.58} & \multicolumn{1}{c}{0.20} & \multicolumn{1}{c}{0.22} \\
\hline
 & \textbf{1: Traditional History} (0.34) & 1: Mathematics (0.28) & 1: Life Science (0.26)\\
Top-3 entities & 2: Astronomy (0.30) & 2: Medicine (0.22) & 2: Law making (0.25)\\
 & 3: Interpretation (0.29) & 3: Astronomy (0.14) & 3: Fine Arts (0.23)\\
\hline
\multicolumn{4}{c}{Result: \textbf{Traditional History}}\\
\bottomrule
\end{tabular}
}
\label{table:case1}
\vspace{-0.4cm}
\end{table}

\begin{table}[t]
\centering
\caption{Case study on RELATION}
\vspace{-0.3cm}
\renewcommand{\arraystretch}{1.1}
\resizebox{\columnwidth}{!}{
\begin{tabular}{llll} 
\toprule
\multicolumn{4}{c}{\textbf{Case 2:} Query (The Lion King, release\_region, ?)} \\
\hline
Semantic facet & \multicolumn{1}{c}{Thematic Genre} & \multicolumn{1}{c}{Actors} & \multicolumn{1}{c}{Market} \\
\hline
Importance & \multicolumn{1}{c}{0.15} & \multicolumn{1}{c}{0.15} & \multicolumn{1}{c}{0.70} \\
\hline
& 1: Army of Palau (-0.6) & 1: Dutch Americans (0.68) & \textbf{1: Argentina} (0.63) \\
Top-3 entities & 2: Matahau, Tonga (-0.7) & 2: Game industry (0.66) & 2: Italy (0.55) \\
& 3: Name of Honduras (-0.78) & 3: Artistic gymnast (0.64) & 3: Denmark (0.54) \\
\hline
\multicolumn{4}{c}{Result: \textbf{Argentina}}\\
\bottomrule
\end{tabular}
}
\label{table:case2}
\vspace{-0.5cm}
\end{table}

\vspace{-0.1cm}
\subsection{Case Study}
\label{sec:case}
\noindent\textbf{Detailed prediction results for two cases.} We conduct a case study to show how MF-CKGE suppresses irrelevant noise and enhances prediction. Table \ref{table:case1} shows a query \textit{(Harvard School of Law, field\_of\_study, ?)} from $Q_1$ of ENTITY after learning $S_3$. For the head entity \textit{Harvard School of Law}, it accumulates three facets over time: \textit{Academic Fields} ($S_1$), \textit{School Symbols} ($S_2$), and \textit{Graduates} ($S_3$). MF-CKGE computes the semantic-aware importance for each facet w.r.t. the relation \textit{field\_of\_study}. The \textit{Academic Fields} facet--most relevant to the query--is assigned the highest importance of 0.58. As a result, the ground-truth entity \textit{Traditional History} is returned as its highest aggregated prediction score. Similarly, Table \ref{table:case2} presents a query \textit{(The Lion King, release\_region, ?)} from $Q_2$ of RELATION after learning snapshot $S_3$. The head entity \textit{The Lion King} accumulates three facets over time: \textit{Thematic Genre} ($S_1$), \textit{Actors} ($S_2$), and \textit{Market} ($S_3$). MF-CKGE assigns the dominant importance (0.7) to the most relevant \textit{Market} facet. Consequently, the ground-truth entity \textit{Argentina} is correctly returned as the final result.

\vspace{0.1cm}

\noindent\textbf{Adaptive importance visualization.} To understand why MF-CKGE correctly answers the above queries, we visualize the average importance assigned to each snapshot embedding $\boldsymbol{E}_i$ across all test sets $Q_j$. A consistent observation across the heatmaps is that the diagonal entries are significantly higher than the off-diagonal ones. This reveals that MF-CKGE can accurately prioritize the most relevant embedding $\boldsymbol{E}_i$ when addressing queries from its corresponding test set $Q_i$. Specifically, on the ENTITY and RELATION datasets, $\boldsymbol{E}_i$ strictly dominates on $Q_i$. For instance, in ENTITY's $Q_5$ and RELATION's $Q_2$, $\boldsymbol{E}_5$ and $\boldsymbol{E}_2$ are assigned exceptionally high weights of 0.53 and 0.49 respectively, whereas the importance of irrelevant embeddings is effectively suppressed (mostly below 0.2). 
Similarly, both the FACT and HYBRID datasets follow this fundamental trend; whether exhibiting distinct local peaks (e.g., 0.42 in HYBRID's $Q_4$) or a relatively smoother attention distribution (as in FACT), their diagonal entries consistently remain the highest within their respective columns. 
Because $Q_i$ is inherently intended to evaluate the knowledge specific to snapshot $S_i$, these results strongly confirm the model's semantic-aware capability. It adaptively identifies the relevance of each embedding to the given query, leveraging critical historical knowledge while filtering out noise from irrelevant snapshots to preserve prediction accuracy.

\begin{figure}[t]
\vspace{-0.2cm}
    \centering
    \includegraphics[width=0.99\linewidth]{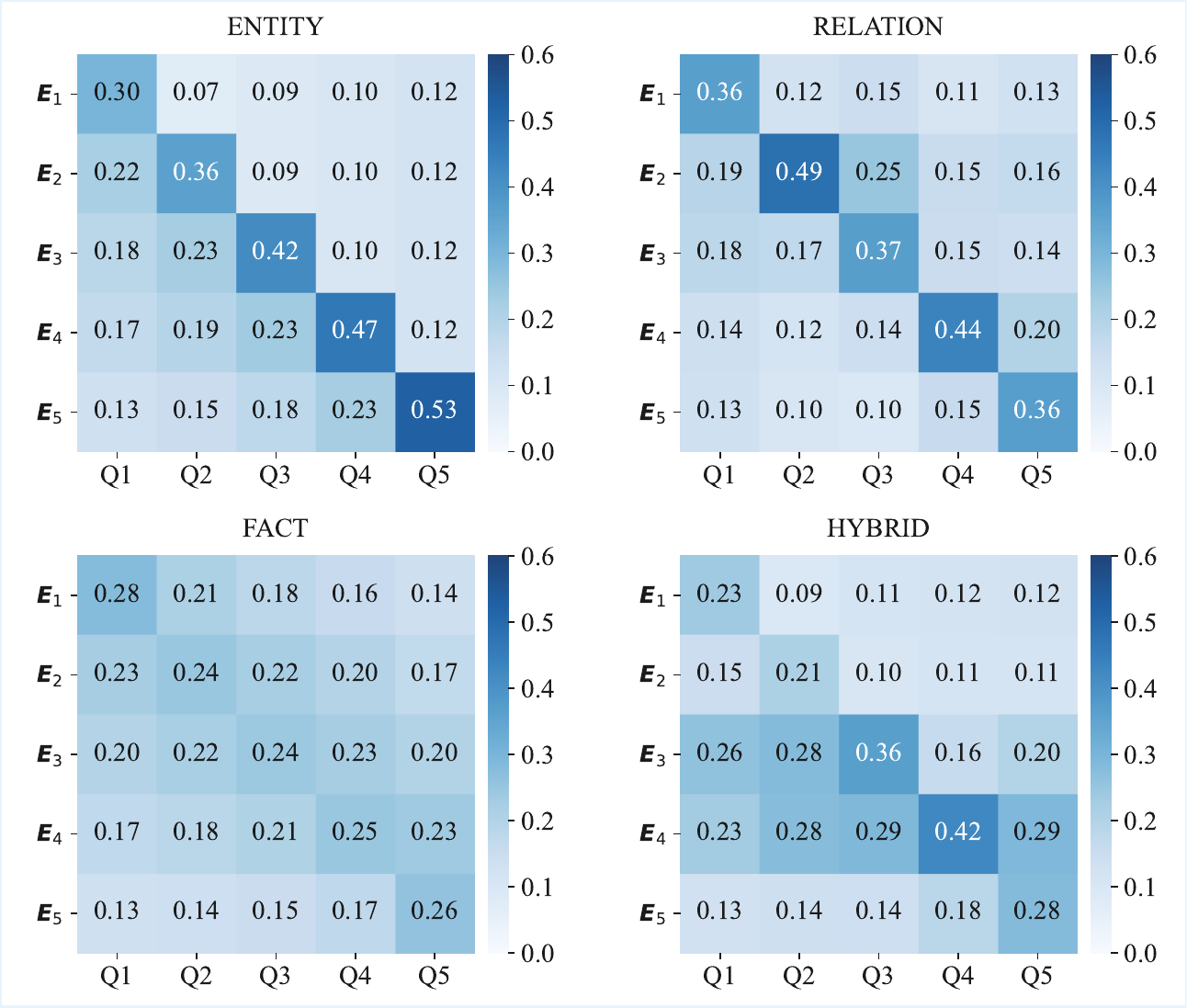}
    \vspace{-0.4cm}
    \caption{Importance for each snapshot embedding}
    \vspace{-0.6cm}
    \label{fig:hotmap}
\end{figure}

\section{Conclusion and Future Work}
\label{sect:conclusion}
This paper proposes an MF-CKGE framework for semantic-aware link prediction. It achieves the separation of new and old knowledge by temporal entity embeddings decoupling, enabling effective learning of new knowledge and retention of old knowledge, and further reduces semantic redundancy via semantic decoupling. We also introduce a semantic-aware importance mechanism that adaptively weights query-relevant snapshot embeddings to improve prediction accuracy. In future work, we plan to build upon this approach to develop a CKGE-enhanced LLM-based RAG system capable of perceiving and adapting to dynamic knowledge changes.


\begin{acks}
This work was supported by the NSFC (62572162 and 62376058), the Primary R\&D Plan of Zhejiang (2023C03198), the "Pioneer" and "Leading Goose" R\&D Program of Zhejiang (No. 2024C01020), and the National Key R\&D Program of China (SQ2025YFE0203954).
\end{acks}

\bibliographystyle{ACM-Reference-Format}
\balance
\bibliography{sample-base}










\end{document}